\documentstyle[mncite,psfig]{mn}

\title[The old nova BT Mon]{The mass of the white dwarf in the old nova BT Mon}

\author[D.\,A.\ Smith et al.]{D.\,A.\ Smith,$^{1}$ 
V.\,S.\ Dhillon,$^2$ T.\,R.\ Marsh$^{3}$ \\
$^1$Institute of Astronomy, Madingley Road, Cambridge CB3 0HA (dsmith@ast.cam.ac.uk) \\
$^2$Royal Greenwich Observatory, Madingley Road, Cambridge CB3 0EZ (vsd@ast.cam.ac.uk)\\
$^3$University of Southampton, Department of Physics and Astronomy, Highfield, 
Southampton SO17 1BJ (trm@astro.soton.ac.uk)}

\date{\small Accepted for publication in the Monthly Notices of the
Royal Astronomical Society, 11 September 1997} 

\begin{document} 
\maketitle

\begin{abstract} 
We present spectrophotometry of the eclipsing old nova BT~Mon (Nova Mon~1939). 
By detecting weak absorption features from the secondary star, we find
its radial velocity semi-amplitude to be $K_R = 205\pm5$\,km\,s$^{-1}$ and 
its rotational velocity to be $v\sin i = 138\pm5$\,km\,s$^{-1}$.
We also measure the radial velocity semi-amplitude of the primary star to
be $K_R = 170\pm10$\,km\,s$^{-1}$.
From these parameters we obtain a mass of $1.04\pm$0.06\,M$_{\odot}$ for the 
white dwarf primary star and a mass of $0.87\pm$0.06\,M$_{\odot}$ for the 
G8\,V secondary star. The inclination of the system is found to be  
$82.2\pm3.2^\circ$ and we estimate that the system lies at a distance 
of $1700\pm300$\,pc. The high mass of the white dwarf and our finding that 
BT~Mon was probably a fast nova together constitute a new piece of evidence 
in favour of the thermonuclear runaway model of classical nova outbursts.  
The emission lines are single peaked throughout the orbital cycle,
showing absorption around phase 0.5, high velocity S-wave 
components and large phase offsets in their radial velocity curves. 
In each of these respects, BT Mon is similar to the SW~Sex stars. We also find 
quasi-periodic flaring in the trailed spectra, which makes BT~Mon a 
candidate intermediate polar.
\end{abstract} 


\section{Introduction}
\label{sec:introduction}

Old novae are cataclysmic binaries in which a white dwarf primary star accretes 
material from a red dwarf secondary star via an accretion disc or magnetic
accretion stream. By definition, an old nova is an object which has been 
observed to have undergone a single nova outburst and then faded back to
its pre-eruptive luminosity (see \pcite{warner95a} for a review).

The thermonuclear runaway model (TNR) is the widely accepted theory of
nova outbursts (see \pcite {starrfield89}). In this model, the primary star builds
up a layer of material, accreted from the secondary star, on its surface. 
The temperature and pressure at the base of the layer eventually become
sufficiently high for nuclear reactions to begin. After ignition, the
temperature rises rapidly and the reaction rates run away, to be stopped only when
the radiation pressure becomes such that most of the envelope is blown away.
The model predicts that the the fastest and most violent TNRs occur as a
result of the high compression in the gravitational wells of the most massive
white dwarfs, while slower TNRs occur on lower mass white dwarfs.
As one of the primary parameters in the TNR model is the 
mass of the white dwarf, we are motivated to try to measure the masses of 
novae as a test. 

The difficulties of measuring the masses of the components of cataclysmic
variables are well known: they can only be measured in eclipsing systems, 
and to date only one old nova, DQ~Her, has had its mass tightly constrained 
(\pcite{horne93} -- $M_1=0.60\pm0.07$M$_\odot$). This was possible because the 
M3\,V secondary star in DQ~Her is 
easily observed through the Na\,{\small I} doublet around $\lambda$8190\AA. DQ~Her 
was a moderately slow nova taking 67 days to fall 2 magnitudes from its
peak luminosity, so its relatively low mass ties in reasonably well with the 
speed of the nova, according to the TNR model.

Nova BT~Mon 1939 is another bright ($V=15.8$), eclipsing old nova and
hence another obvious candidate for mass determination. 
It has a longer period than DQ~Her so the secondary star is of an earlier 
spectral type, and because the accretion structures are bright, the secondary star 
is faint compared with the system as a whole, so we are
forced to adopt different observation and analysis strategies to those
of Horne et al. (1993), as described below. 

\section{Observations}
\label{sec:observations} 

On the night of 1995 January 23/24 we obtained 47 red and 49 blue spectra of the 
old nova BT~Mon, covering $\sim1.0$ orbits from cycle 18718.87 to 18719.86 
(according to our ephemeris, see section 4.1) with
the 4.2-m William Herschel Telescope (WHT) on La Palma. 
The exposures were all around 600-s with about 15-s dead-time for the dumping of 
data.
The ISIS spectrometer with the R1200R and R600B gratings 
and the TEK CCD chips gave a wavelength coverage of approximately $4570-5370$\AA\
at $0.8$\AA\ (50\,km\,s$^{-1}$) resolution in the blue arm, and $6310-6720$\AA\
at $0.4$\AA\ (18\,km\,s$^{-1}$) resolution in the red arm. 
We also took spectra of the spectral type templates 61~UMa (G8\,V), 
Gleise~567 (K0.5\,V), Gleise~28 (K2\,V), Gleise~105 (K3\,V),
Gleise~69 (K5\,V), EQ~Vir (K5\,V) and Gleise~380 (K6\,V).
The 1 arcsecond slit was oriented to cover a nearby field star in order to 
correct for slit losses.
Comparison arc spectra were taken every $30-40$ min to calibrate
flexure. The night was photometric and the seeing ranged from 1.5 to 3 arcseconds.

\section{Data Reduction} 
\label{sec:datared} 
We first corrected for 
pixel-to-pixel variations with a tungsten lamp flat-field.
After sky subtraction, the data were optimally extracted
to give raw spectra of BT Mon and the comparison star.
Arc spectra were then extracted from the same locations 
on the detector as the targets. The wavelength 
scale for each spectrum was interpolated from the wavelength scales of two
neighbouring arc spectra. The spectra of the nearby star were used to correct for 
slit losses and the observations were placed on an absolute flux scale by using 
observations of the standard star Feige~34 \cite{oke90} taken immediately after the run.

\section{Results}
\label{sec:results}
\subsection{Ephemeris}
\label{sec:ephemeris}

The time of mid-eclipse for our observation was determined by plotting the 
blue continuum light curve (see figure~2)
about the eclipse (following \pcite{robinson82}, hereafter RNK, whose measurements
were approximately in the blue), and then fitting the midpoints of chords.
A linear least-squares fit to the eight eclipse timings of RNK, the timing of 
\scite{seitter84}, and our eclipse timing (all of which are 
presented in table~1) yield the following ephemeris:  

\begin{equation}
\begin{array}{rrrl}
T_{\rm mid-eclipse} = & \!\!\!\!\! {\em HJD}\,\,2\,443\,491.7159 
& \!\! + \,\, 0.33381379 & \!\!\!\!\! E \\
& \!\! \pm \,\, 0.0001 & \!\! \pm \,\, 0.00000001 & \\
\end{array}
\end{equation}

We find no evidence of a non-zero value for $\dot P$, in agreement with \scite{seitter84}.

\begin{table*}
\caption{Times of mid-eclipse for BT~Mon. $^1$ 
Seitter's uncertainty is taken to be the same as ours.} 

\boldmath{
{\normalsize\bf
\begin{tabular}{lcrc} 
 & & \\
\multicolumn{1}{l}{Cycle} &
\multicolumn{1}{c}{HJD at mid-eclipse} & 
\multicolumn{1}{c}{O--C} &
\multicolumn{1}{c}{Reference} \\
\multicolumn{1}{c}{(E)} & 
\multicolumn{1}{c}{(2,440,000+)} &
\multicolumn{1}{c}{(secs)} &
\multicolumn{1}{c}{ } \\
 & & \\
0	& $3491.7168\pm0.0020$ & 81.3 & RNK \\
3	& $3492.7144\pm0.0006$ & --250.6 & RNK \\
81	& $3518.7540\pm0.0002$ & --67.0 & RNK \\
3176	& $4551.9097\pm0.0002$ & 107.0 & RNK \\
3179	& $4552.9098\pm0.0002$ & --8.8 & RNK \\
3182	& $4553.9125\pm0.0002$ & 99.9 & RNK \\
3188	& $4555.9134\pm0.0002$ & --71.4 & RNK \\
3433	& $4637.6971\pm0.0002$ & --130.1 & RNK \\
5488    & $5323.6875\pm0.0002$ & 134.2 & Seitter$^1$ \\
18719	& $9740.3758\pm0.0002$ & --36.7 & This paper \\
 & & &\\
\end{tabular}
}
}
\label{tab:ephemeris}
\end{table*}

\subsection{Average spectrum}
\label{sec:average}

\begin{table*}
\caption{Fluxes and widths of prominent lines in BT~Mon, measured
from the average spectrum.}
\boldmath{
{\normalsize\bf
\begin{tabular}{lcrrcc} 
 & & & & & \\
\multicolumn{1}{l}{Line} &
\multicolumn{1}{c}{Flux} & 
\multicolumn{1}{c}{EW} & 
\multicolumn{1}{c}{FWHM} & 
\multicolumn{1}{c}{FWZI} \\
\multicolumn{1}{l}{ } &
\multicolumn{1}{c}{$\times$ 10$^{-14}$} &
\multicolumn{1}{c}{\AA} & 
\multicolumn{1}{c}{km\,s$^{-1}$} & 
\multicolumn{1}{c}{km\,s$^{-1}$} \\
 & erg\,cm$^{-2}$\,s$^{-1}$ & & & & \\
& & & & &\\
H$\alpha$ & $3.9\pm0.1$ & $28.3\pm0.1$ & \ $950\pm100$ & $4500\pm500$ \\
H$\beta$  & $2.4\pm0.1$ & $13.1\pm0.1$ & \ $750\pm100$ & $3500\pm500$ \\
He\,{\small\bf I} $\lambda$6678\AA\  & $0.3\pm0.1$ & \ $2.5\pm0.1$ & \ $600\pm10$0 & $1500\pm500$ \\
He\,{\small\bf II} $\lambda$4686\AA\ & $4.2\pm0.1$ & $21.4\pm0.1$ & \ $750\pm100$ & $4500\pm500$ \\
C\,{\small\bf III}/N\,{\small\bf III} $\lambda\lambda$4640-4650\AA\ & $1.6\pm0.1$ & \ $8.5\pm0.1$ & $1000\pm100$ & $3500\pm500$ \\
& & & & & \\
\end{tabular}
}
}
\label{tab:linewidths}
\end{table*}
\rm
The average spectrum of BT~Mon is displayed in figure~1,
and in table~2 we list fluxes, equivalent widths and 
velocity widths of the most prominent lines measured from the average 
spectrum.

BT~Mon shows broad Balmer and He\,{\small I} lines and high excitation lines of 
He\,{\small II} $\lambda$4686\AA\ and 
C\,{\small III}/N\,{\small III} $\lambda\lambda$4640-4650\AA. 
The spectrum has similarities with those of other high inclination
nova-likes e.g. SW~Sex \cite{thorstensen91} and
intermediate polars e.g. FO~Aqr \cite{marsh96} 
in that the emission lines do not show the double peaks 
characteristic of high-inclination accretion discs \cite{horne86}.
The weak, narrow 
absorption features around $\lambda$6380\AA\ and $\lambda$6620\AA\ show no 
radial velocity variations and can be attributed 
to the interstellar medium (as BT~Mon lies at a galactic latitude of $-3\degr$
and at a distance of $\sim 2$\,kpc). There is evidence for faint absorption 
features from the secondary star at $\lambda$5170\AA\ and $\lambda$6500\AA.

\begin{figure*}
\centerline{\psfig{file=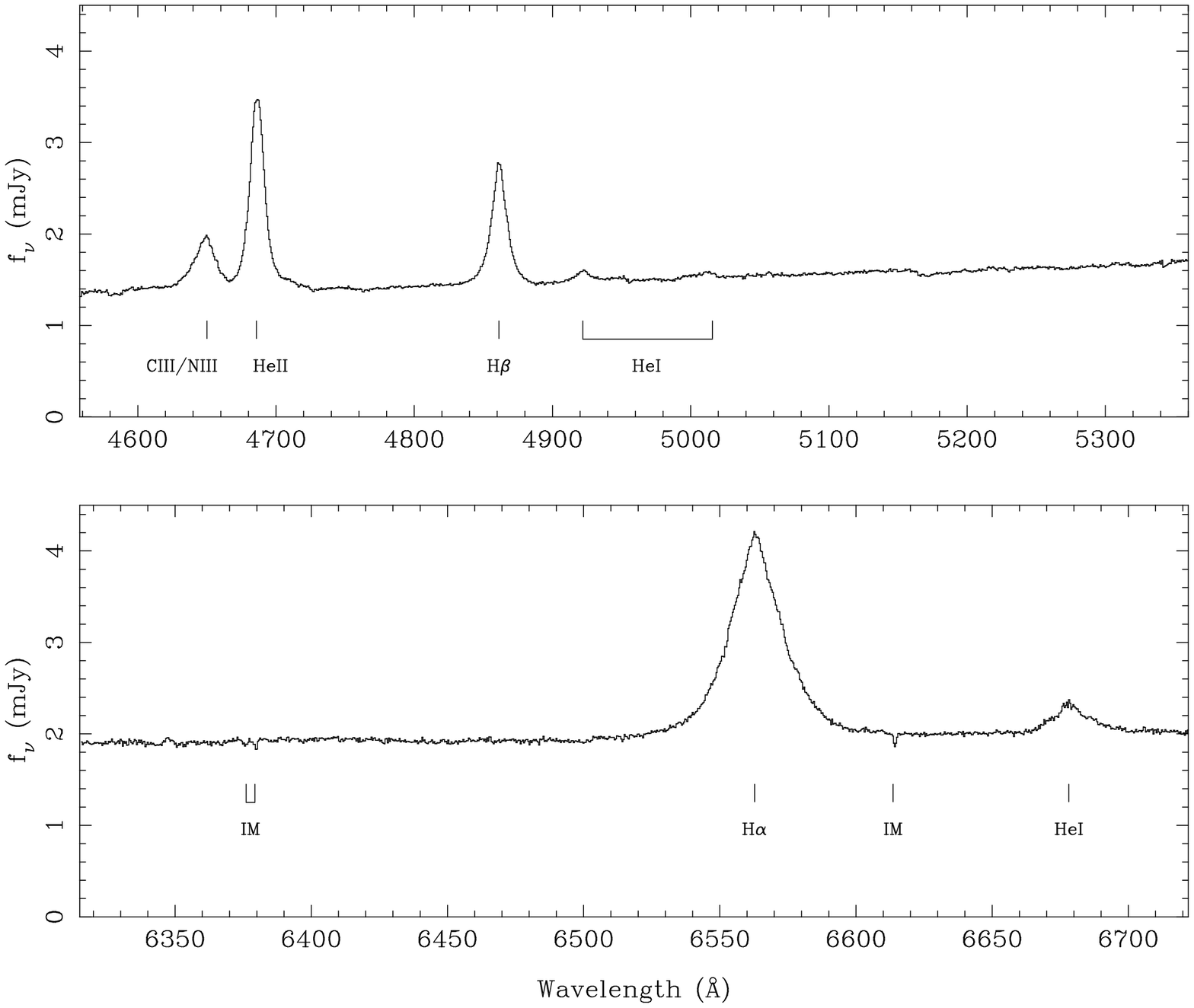,height=130mm,angle=-90}}
\caption{The average spectrum of BT~Mon, uncorrected for 
orbital motion.}
\label{fig:average}
\end{figure*}

\subsection{Light curves}
\label{sec:light}

Regions of the spectrum devoid of emission lines were selected
in the red and blue ($\lambda\lambda$6390--6490\AA\ and 
$\lambda\lambda$4730--4830\AA). The red and blue continuum light curves
for BT~Mon were then computed by summing the flux in the above wavelength ranges. 
A third order polynomial fit to the continuum was subtracted 
from the red and blue spectra and the emission-line light curves were then computed 
by summing the residual flux between $\pm2000$\,km\,s$^{-1}$ for 
the Balmer lines, $\pm1000$\,km\,s$^{-1}$ for He\,{\small I} $\lambda$6678\AA\ 
and between $\lambda\lambda$4620-4730\AA\ for He\,{\small II} and 
C\,{\small III}/N\,{\small III}.

The resulting light curves are plotted in figure~2 as a function of phase,
following our new ephemeris. The continuum shows a deep symmetrical eclipse,
the blue light curve having a deeper eclipse than the red. There is no sign 
of an orbital hump at $\phi\sim0.9$, as seen by \scite{seitter84}, but 
there is flickering throughout.
The continuum light curves of BT~Mon in the red and the blue are similar 
to those of the well-known eclipsing cataclysmic variables DQ~Her, RW~Tri 
and UX~UMa, as noted by RNK.  

The eclipses of the Balmer lines have a different shape to the continuum, 
there being a distinctive shoulder in the light curve as it enters eclipse.
A similar shoulder is seen in the light curve of the weak 
He\,{\small I} line, but the light curve of He\,{\small II} shows only a hint of it.
The lines are all deeply eclipsed, even appearing flat-bottomed in the Balmer
and He\,{\small II} lines, with the eclipse in H$\alpha$ being shallower than 
in the other lines, which almost vanish completely. There is flickering in all 
of the emission line light curves, and a significant drop in the
Balmer and He\,{\small I} line flux at $\phi\sim0.5$.
The emission line light curves of BT~Mon are similar 
to those of SW~Sex \cite{dhillon97} and the other systems in the SW~Sex class, 
especially with regard to the phase 0.5 absorption.
 
\begin{figure*}
\label{fig:lc}
\centerline{\psfig{file=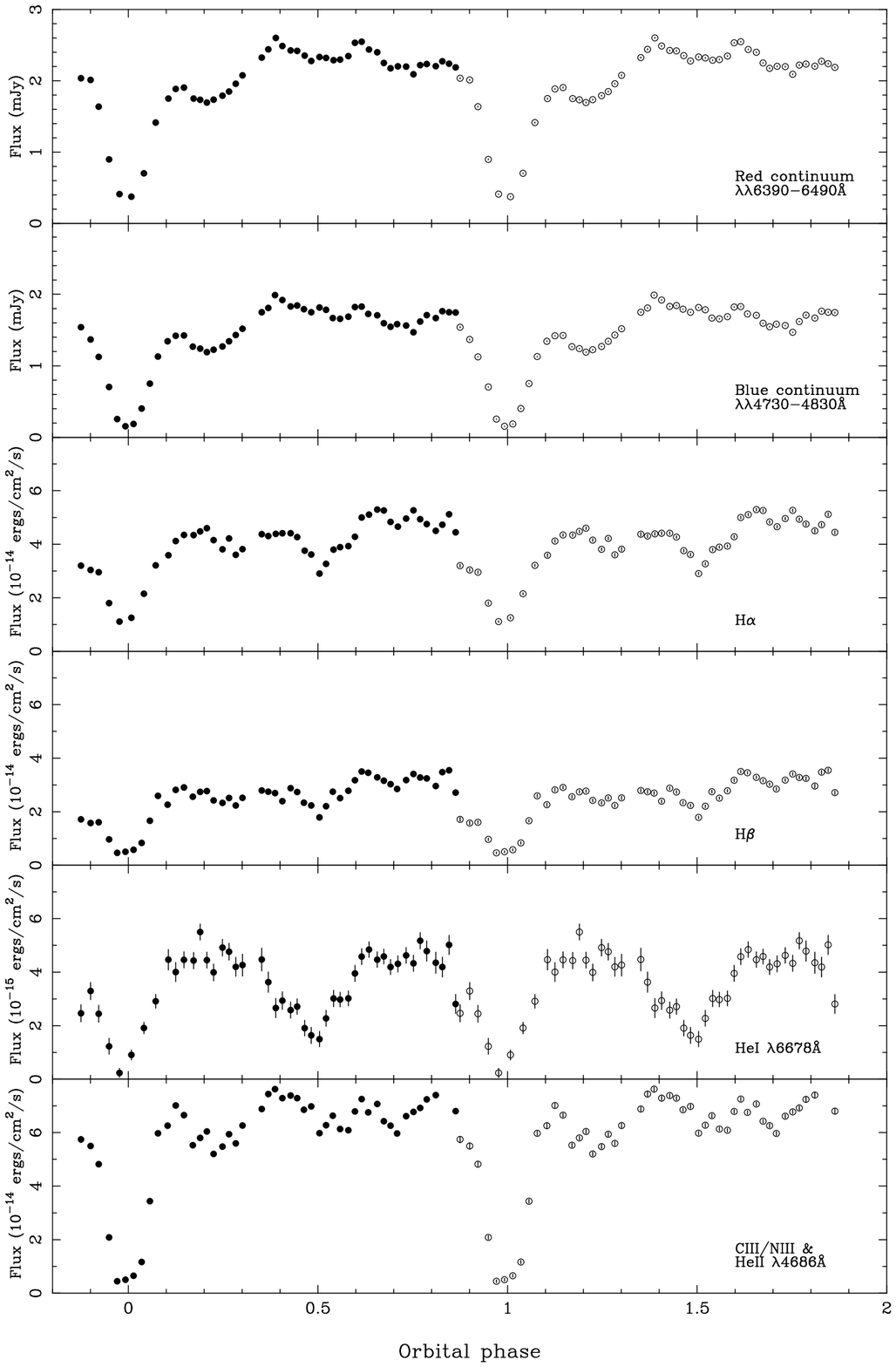,height=190mm}}
\caption{Continuum and emission-line light curves of BT~Mon. The open circles represent 
points where the real data, the closed circles, have been folded over.}
\end{figure*}

\subsection{Line profile evolution}
\label{sec:evolution}

We rebinned all of the spectra onto a uniform wavelength scale, and
placed the data into ten binary phase bins by averaging all the spectra 
falling into each bin. A multiple of 2.8 was then added to each spectrum 
in order to displace the data in the $y$-direction. The result is plotted 
in figure~3.

The Balmer line profiles vary dramatically over one orbital cycle. 
The H$\alpha$ line grows from a small bump at $\phi\sim0.0$ to a narrow 
spike at $\phi\sim0.2$, then evolves into a broader, multi-peaked 
hump at $\phi\sim0.5$, rising again to a spike at $\phi\sim0.7$,
mirroring that at $\phi\sim0.2$, 
and finally declining on the way to eclipse. H$\beta$ and He\,{\small I} undergo 
similar changes, but they seem to be single-peaked throughout, although
this could be an artifact of the lower resolution.
They become narrow  
at $\phi\sim0.2$ and $\phi\sim0.7$ and broader at $\phi\sim0.5$, vanishing 
altogether at $\phi\sim0.0$. The
high excitation lines of He\,{\small II} and C\,{\small III}/N\,{\small III} 
vary little over one binary orbit,
except for their total disappearance during eclipse, and seem to
be single-peaked at all phases.

\begin{figure*}
\label{fig:evol}
\centerline{\psfig{file=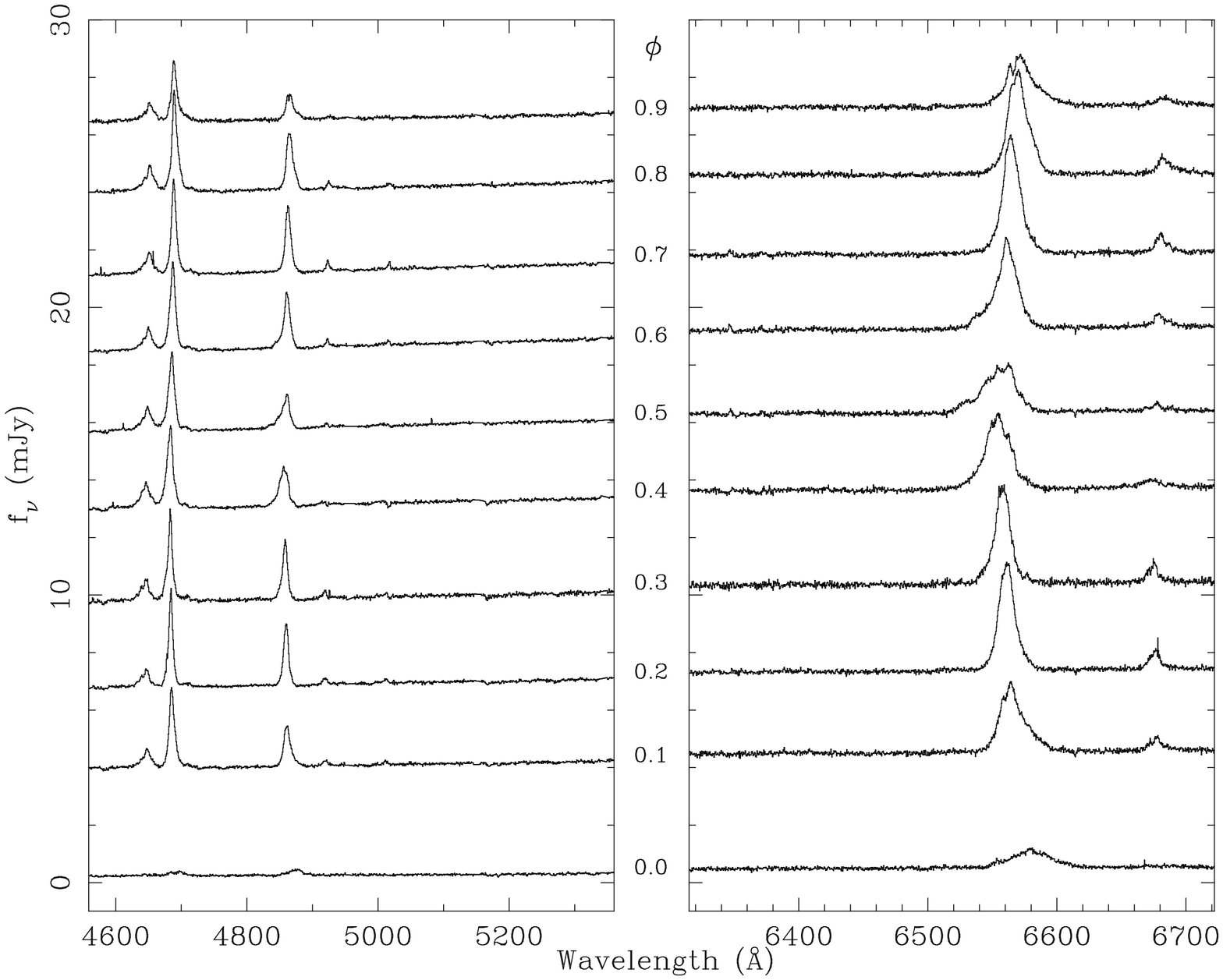,height=5.5in,angle=-90}}
\caption{Orbital emission line variations in BT~Mon. The data have been 
averaged into 10 binary phase bins with a multiple of 2.8 added to each 
spectrum in order to displace the data in the $y$-direction.}
\end{figure*}

\subsection{Trailed spectra}
\label{sec:trail}

We subtracted the continua from the spectra using a third order polynomial fit and then 
rebinned the spectra onto a constant velocity interval scale centred on the rest wavelengths 
of the lines. The upper panels of figure~4 show the trailed spectra of
the H$\alpha$, H$\beta$, He\,{\small II} $\lambda$4686\AA\ and
He\,{\small I} $\lambda$6678\AA\ lines in BT~Mon. Each shows the characteristic 
sine waves of orbital motion in the core of the emission line, 
but in addition to this, there are clear signs of high velocity Balmer emission 
moving from $\sim2000$\,km\,s$^{-1}$ at $\phi\sim0$ to $\sim-2000$\,km\,s$^{-1}$ at 
$\phi\sim0.5$.
The high velocity emission is also present in the He\,{\small II} line but is weaker. 
Looking at figure~4 one notices that the semi-amplitudes of the emission line cores
vary, with H$\alpha$ having the highest velocity semi-amplitude, H$\beta$ 
the next highest, and He\,{\small II} the lowest.
This agrees with the observations of \scite{seitter84} 
and \scite{white96}. Note also the periodic flaring in the trailed spectra,
visible in the horizontally striped appearance of the upper panels.
This will be discussed in more detail in section~\ref{sec:period}.
The cores of the Balmer lines almost disappear due to 
the phase 0.5 absorption, and the eclipse shows nothing
other than the high velocity S-wave emission in the Balmer lines, which 
is never totally eclipsed. There are also 
faint components running vertically through the trailed Balmer-line spectra,
which could be lines from the nebula \cite{marsh83}.

\subsection{Doppler tomography}
\label{sec:doppler}

Doppler tomography is an indirect imaging technique which can be used to 
determine the velocity-space distribution of line emission in cataclysmic
variables. In this study we used the maximum entropy method to create the 
Doppler maps. Full technical details of the method, including a number of 
test simulations are given by \scite{marsh88b}.
Examples of the application of Doppler tomography to real data are given 
by \scite{marsh90b} and \scite{marsh90c}. 

Figure~5 shows the Doppler maps of the H$\alpha$, H$\beta$, 
He\,{\small I} $\lambda$6678\AA\ and He\,{\small II} $\lambda$4686\AA\ lines 
in BT~Mon, computed from the trailed spectra of figure~4 
but with the eclipse spectra removed. 
The three crosses on the Doppler maps represent the centre of mass of the 
secondary star (upper cross), the centre of mass of the system (middle cross)
and the centre of mass of the white dwarf (lower cross). The secondary star's
Roche lobe and the predicted trajectory of the gas stream have been plotted
using the mass ratio, $q=M_2/M_1=0.84$ derived in section~\ref{sec:parameters}. 
The series of circles 
along the path of the gas stream mark the distance from the white dwarf at 
intervals of $0.1L_1$, ranging from $1.0L_1$ at the red star to $0.1L_1$ 
at the point of closest approach, marked by an asterisk.

It is clear from the Doppler maps that the emission regions do not coincide
with either the gas stream or the secondary star. A ring-like emission 
distribution, characteristic of a Keplerian accretion disc about the primary star
is also absent from the maps. 
As in most nova-likes the bulk of the Balmer emission is mapped to the lower 
left quadrant of the tomogram, with the high velocity feature from the trailed
spectra mapped to a zone extending out to $V_x=-2000$\,km\,s$^{-1}$.
The He\,{\small II} emission seems to be 
centred on or near the expected position of the white dwarf.
A likely origin of the line emission is discussed in section~\ref{sec:nature}. 

Trailed spectra have been computed from the Doppler maps and compared to 
the original trailed spectra as a check. They are plotted in the lower panels 
of figure~4 with the eclipse spectra omitted. They compare 
favourably, showing the original S-waves and high velocity features.

\begin{figure*}
\label{fig:trail}
\centerline{\psfig{file=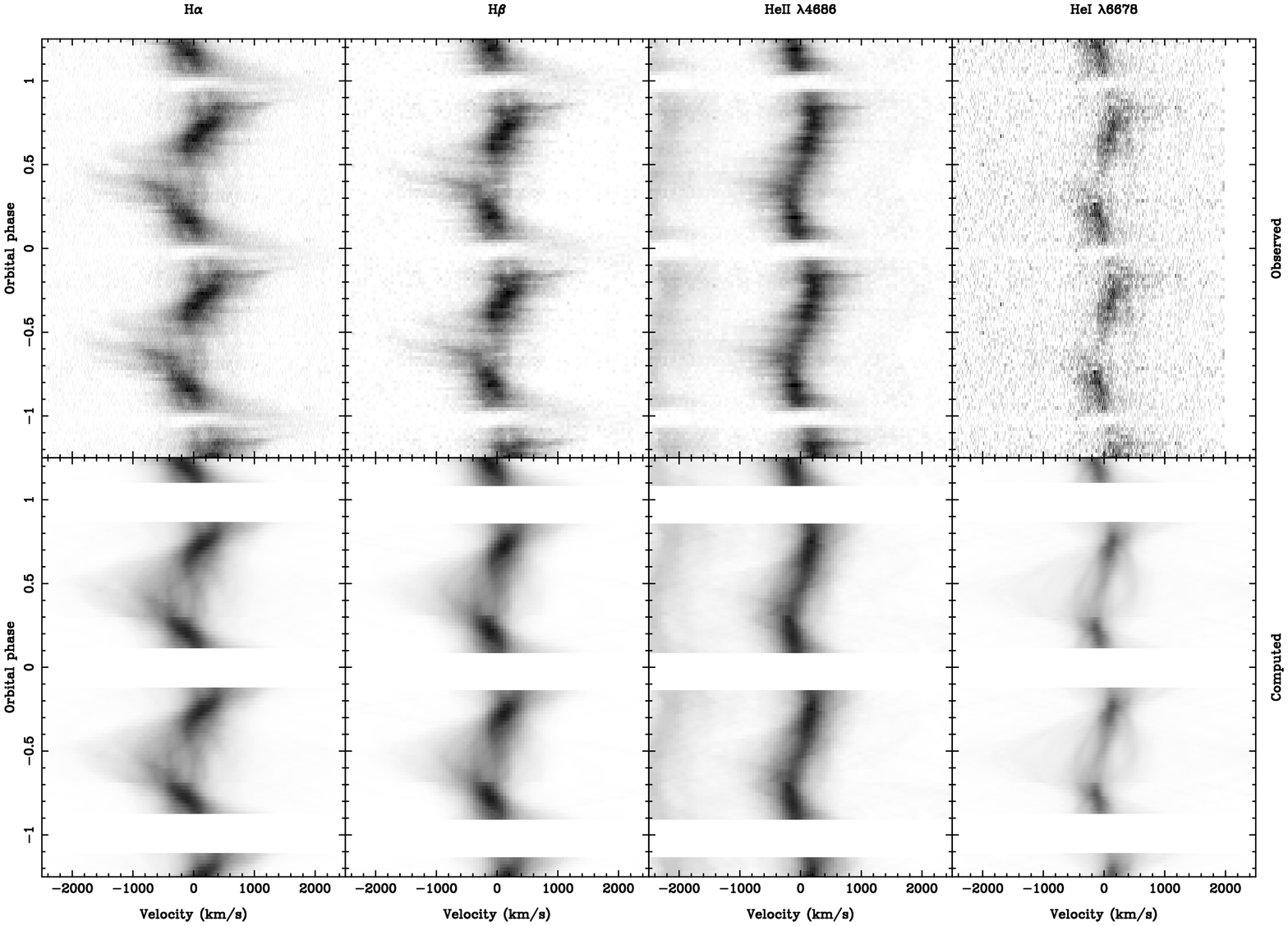,height=5.2in,angle=-90}}
\caption{Trailed spectra of H$\alpha$, H$\beta$, 
He\,{\small II} $\lambda$4686\AA\ and He\,{\small I} $\lambda$6678\AA\ in
BT~Mon are displayed in the upper panels with
the velocity relative to the line centre along the horizontal axis and the
orbital phase along the vertical axis. The lower panels show the fits 
computed from the Doppler maps; the gaps correspond to eclipse data which were 
omitted from the fit. The data cover only one cycle, but are folded over for clarity.}
\end{figure*}

\begin{figure*}
\label{fig:map}
\centerline{\psfig{file=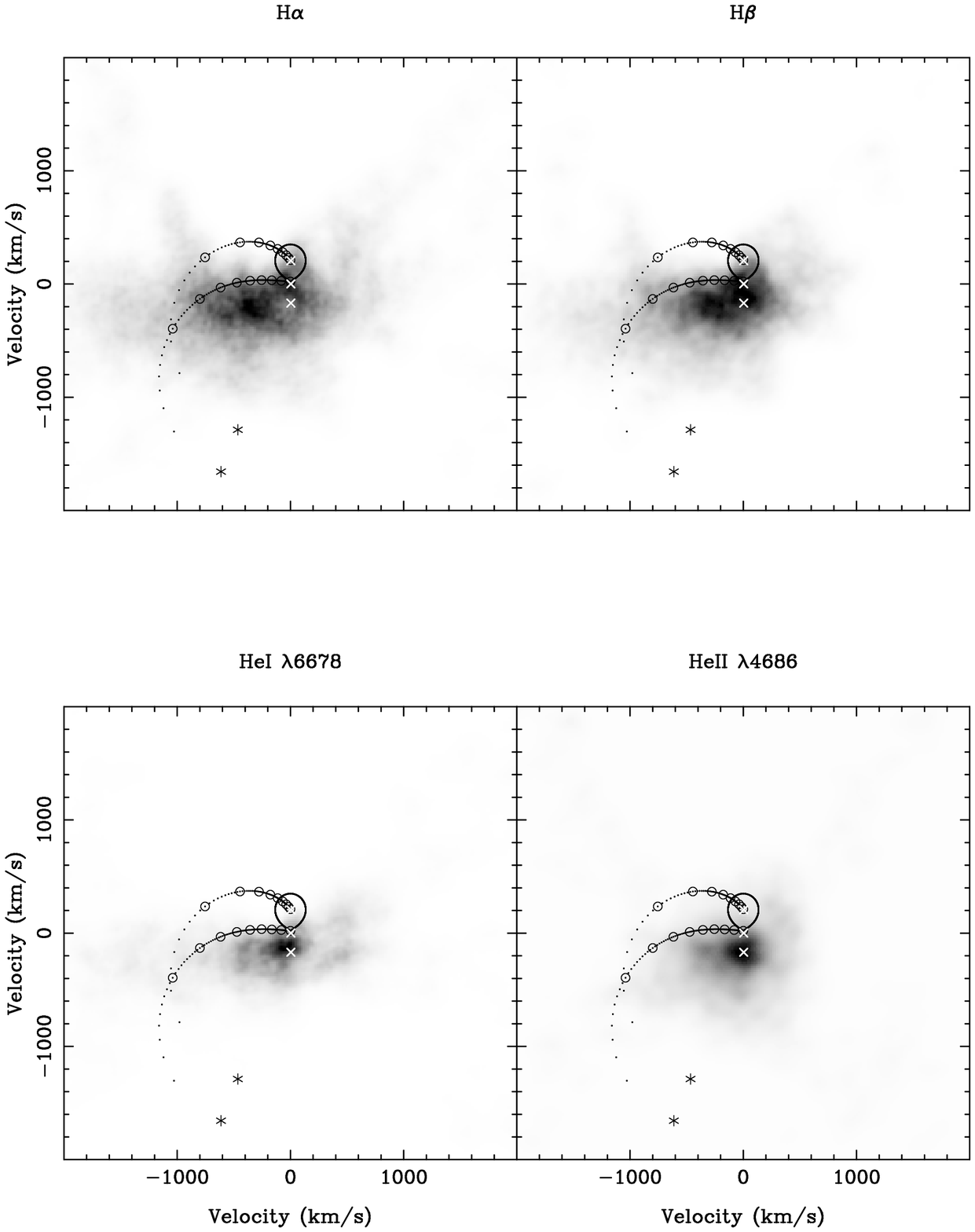,height=7.5in}}
\caption{Doppler maps of BT~Mon in H$\alpha$, H$\beta$, 
He\,{\small I} $\lambda$6678\AA\ and He\,{\small II} $\lambda$4686\AA\ computed 
from the trailed spectra in figure~4. The predicted position of the red star,
the path of the gas stream (the lower curve) and the Keplerian velocity at the gas stream 
(upper curve) are marked. The three crosses on the map are, 
from top to bottom, the centre of mass of the red star, the system (at zero velocity) 
and the white dwarf. The emission at $\sim(-1400,-200)$\,km\,s$^{-1}$ corresponds
to the high velocity S-wave. }
\end{figure*}

\subsection{Flares in the trailed spectra}
\label{sec:period}
The trailed spectra of figure~4 show periodic flaring similar to the features seen 
more clearly in the trailed spectra of intermediate polars e.g. FO~Aqr \cite{marsh96}. 
Because the flares are weak in BT~Mon we attempted to 
enhance them by removing the core of the line emission and the high velocity S-wave.
This was achieved by smoothing the trailed spectra images and subtracting the smoothed 
image from the original. The resulting trailed spectra are displayed in 
the top panel of figure~6 and appear to show a periodicity of $\sim30$~minutes. 

The flare spectra were then straightened in the velocity direction
using the value of the white dwarf radial 
velocity found in section~4.12 ($K_W = 170$\,km\,s$^{-1}$). The shifted
spectra are plotted in the central panels of figure~6.
Light curves at each wavelength were calculated for these straightened flare
spectra and trailed power spectra of the three lines were then obtained by 
applying the
Lomb-Scargle periodogram technique \cite{press89} to each of these light curves. 
The result is plotted in the bottom panels of figure~6. 
H$\alpha$ shows no clear peaks, but
the H$\beta$ and He\,{\small II} $\lambda$4686\AA\ periodograms have
several peaks around 45 cycles per day, suggesting that the flares may
be quasi-periodic. 

The most obvious model for the multiperiodicity is that of the system being an 
intermediate polar, with material channelled down a magnetic accretion stream
onto the magnetic pole of an asynchronously rotating primary star. The polar regions become
hot, emitting X-rays which periodically 
light up the accretion stream/disc/curtain like a lighthouse.
Further evidence, however is required to be certain of this classification
(coherence of pulses, X-ray detection; see section~5.1). 

\begin{figure*}
\centerline{\psfig{file=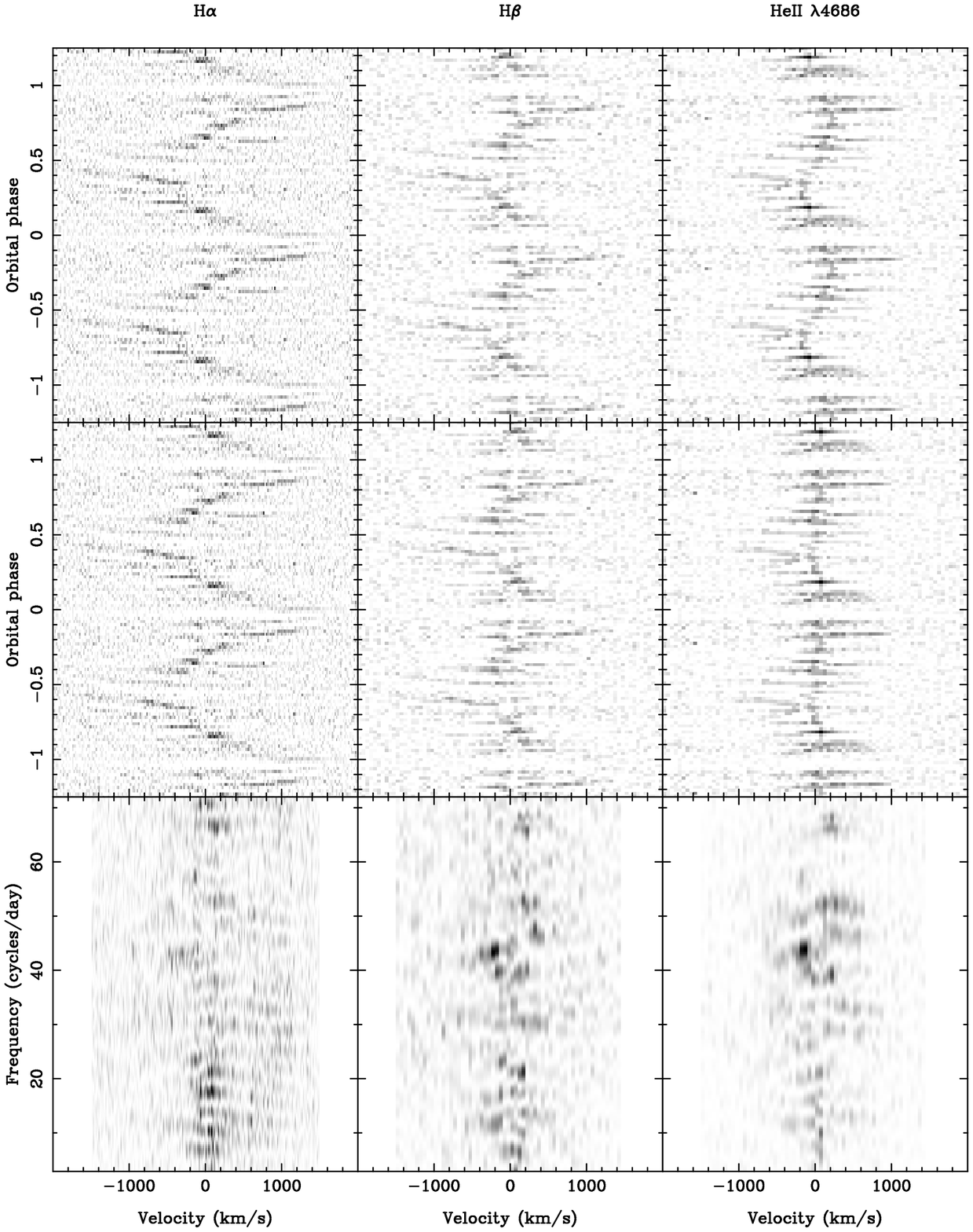,height=6.5in,angle=0}}
\caption{Upper panels: trailed spectra of H$\alpha$, H$\beta$, and 
He\,{\small II} $\lambda$4686\AA\ with the flares enhanced (see text for details).
Central panels: the same as the upper panels, but straightened 
to account for orbital motion.
Lower panels: power spectra computed from the central panels
at each velocity, with frequency running vertically. } 
\label{fig:flares}
\end{figure*}

\subsection{Radial velocity of the white dwarf}
\label{sec:radial}
The continuum-subtracted spectra were binned 
onto a constant velocity interval scale about each of the emission-line 
rest wavelengths. In order to measure the radial velocities, we used the 
double-Gaussian method of \scite{schneider80}, since this technique is
sensitive mainly to the motion of the line wings and should therefore reflect 
the motion of the white dwarf with the highest reliability. The Gaussians 
were of width 200\,km\,s$^{-1}$ (FWHM) and we varied their separation $a$ from 
400 to 3600\,km\,s$^{-1}$. We then fitted 
\begin{equation}
V=\gamma-K\sin(\phi-\phi_0)
\end{equation} 
to each set of measurements, omitting 11 points near primary eclipse. 
Examples of the radial velocity 
curves obtained for H$\alpha$, H$\beta$, He\,{\small I} $\lambda$6678\AA\ and 
He\,{\small II} $\lambda$4686\AA\ 
for Gaussian separations of 1600\,km\,s$^{-1}$ 
are shown in figure~7. 

The radial velocity curves of BT~Mon resemble those of other novae e.g. DQ~Her,
and intermediate polars e.g. FO~Aqr, where the superior 
conjunction of the emission line source, occurs after photometric 
mid-eclipse. This phase shift implies an 
emission-line source trailing the accretion disc, such as the bright-spot,
and has also been observed in many nova-likes including 
V1315~Aql \cite{dhillon91} and SW~Sex (Dhillon et al. 1997).

The Doppler maps and light curves seem to indicate that the 
source of the He\,{\small II} $\lambda$4686\AA\ emission is centred on the white dwarf. 
This would be expected,
since the He\,{\small II} line is a high excitation feature normally only
seen in high temperature regions e.g. near the surface of the white dwarf.
However the radial velocity curves produced by using the double-Gaussian technique
show large phase shifts.
A visual inspection of the trailed spectra seems to show that it is the high 
velocity features in the wings which are phase shifted while the 
core is not. As the double-Gaussian approach to measuring the velocity 
would then be dominated by emission from the high velocity feature,
we instead attempted to measure the radial velocity of the
He\,{\small II} line by fitting the core of the emission line, with a single 
Gaussian with FWHM ranging from 50 to 800\,km\,s$^{-1}$. The radial velocity
curve produced using 
a single Gaussian of width 100\,km\,s$^{-1}$ is also plotted in figure~7
and shows a much lower semi-amplitude and a smaller phase shift, which 
appears to confirm that this emission does indeed come from close to the white dwarf.

The results of the radial velocity analysis are  
displayed in the form of a diagnostic diagram in figure~8.
By plotting $K$, its associated
fractional error $\sigma_K/K$, $\gamma$ and $\phi_0$ as functions of the Gaussian 
separation in the case of the double-Gaussian fits, or Gaussian FWHM in the
case of the single Gaussian fits, it is supposedly possible to select the value 
of $K$ which most closely matches $K_W$ \cite{shafter86}. 
If the emission were disc dominated,
one would expect the solution for $K$ to asymptotically reach $K_W$ when the
Gaussian separation becomes sufficiently large, and furthermore, one would
expect $\phi_0$ to fall to 0. This however is not the case as the phase shift
$\phi_0$ is over 0.1 and increases with larger double Gaussian separation. The 
phase shift of the single
Gaussian fits for He\,{\small II} $\lambda$4686\AA, however, do seem to  
fall towards zero with decreasing FWHM.

We therefore attempted to make use of a modified version of 
the light centres method, as described by 
\scite{marsh88a}. In the co-rotating co-ordinate system, the white dwarf 
has velocity ($0, -K_W$), and symmetric emission, say from a disc, would 
be centred at that point. By plotting $K_x = -K\sin\phi_0$ 
against $K_y = -K\cos\phi_0$ for the different 
radial velocity fits (figure~9), one finds that  
with decreasing values of the FWHM of the core Gaussian fit the points move 
closer to the $K_y$ axis.
This is because narrower Gaussian are less affected by the wings of the 
line and more accurately follow the radial 
velocity of the peak. A linear fit to the points on the light centre diagram 
can be extrapolated to the $K_y$ axis to give a measurement of $K_W$.
The extrapolation to the velocity of the white dwarf is fairly uncertain, but 
in figure~9 we have underplotted the Doppler tomogram. We identify the 
position of the peak intensity in the tomogram with the radial
velocity of the white dwarf, giving a value of $K_W = 170 \pm 10$\,km\,s$^{-1}$.

\begin{figure*}
\centerline{\psfig{file=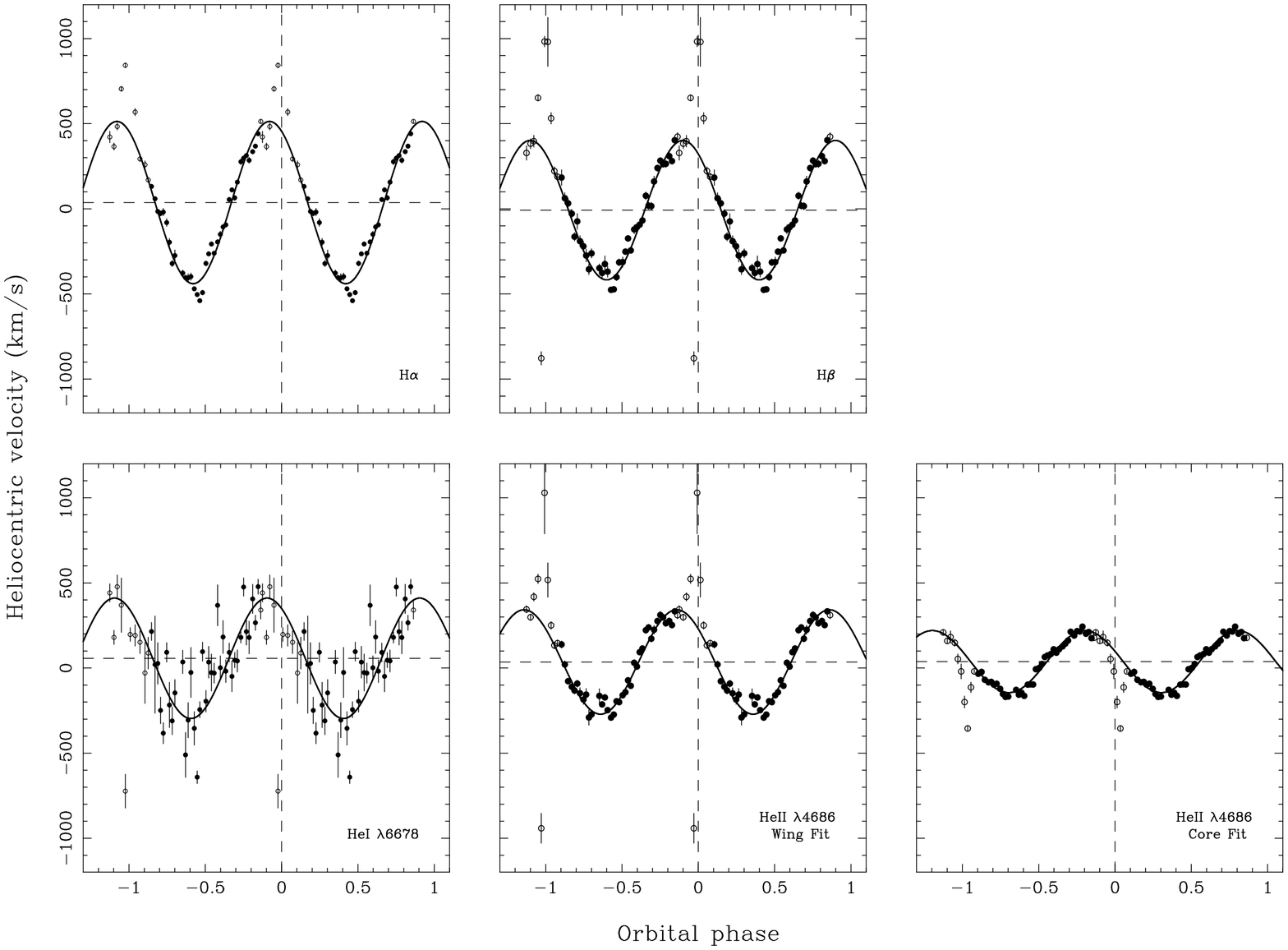,height=130mm,angle=-90}}
\caption{Radial velocity curves of H$\alpha$, H$\beta$, 
He\,{\small I} $\lambda$6678\AA\ and He\,{\small II} $\lambda$4686\AA\ 
measured using a double Gaussian fit with a Gaussian separation of 1600\,km\,s$^{-1}$.
The bottom-right panel shows the radial velocity curve of 
He\,{\small II} $\lambda$4686\AA, measured using a single Gaussian of width 
100\,km\,s$^{-1}$ FWHM. Points marked by open circles were not included in the radial 
velocity fits, due to measurement uncertainties during eclipse. The horizontal 
dashed lines represent the systemic velocities.} 
\label{fig:rvcurve}
\end{figure*}

\begin{figure*}
\centerline{\psfig{file=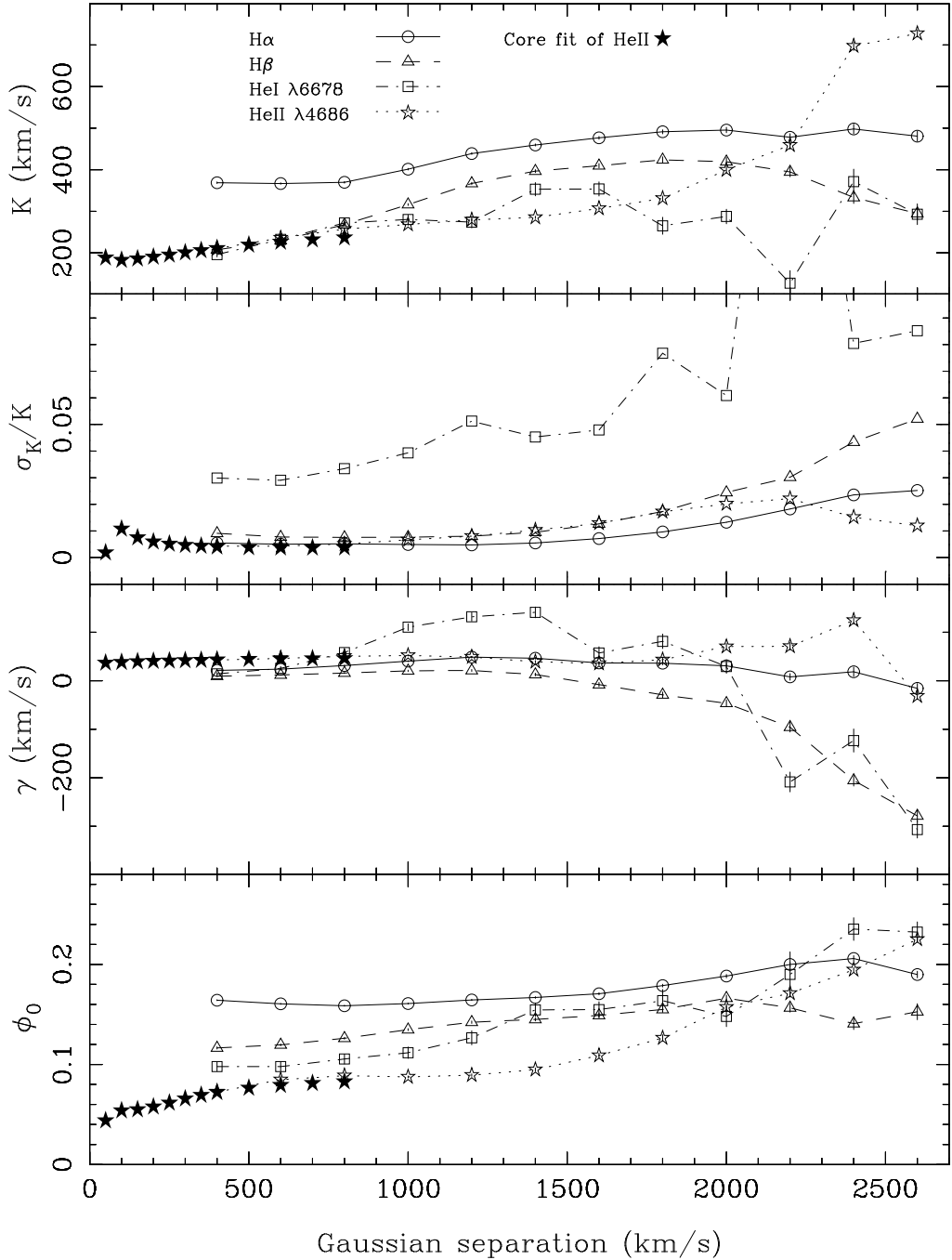,height=180mm,angle=0}}
\caption{The diagnostic diagram for BT~Mon based on the double Gaussian 
radial velocity fits 
to H$\alpha$ (circles connected by solid line), H$\beta$ (triangles connected
by dashed lines), He\,{\small I} $\lambda$6678\AA\ (squares connected by 
dashed-dotted lines) and He\,{\small II} $\lambda$4686\AA\ (stars connected by 
dotted lines). Radial velocity fits for He\,{\small II} $\lambda$4686\AA\ using
a core Gaussian fit are also plotted against the FWHM in km\,s$^{-1}$
with solid stars.}
\label{fig:diagnostic}
\end{figure*}

\begin{figure*}
\centerline{\psfig{file=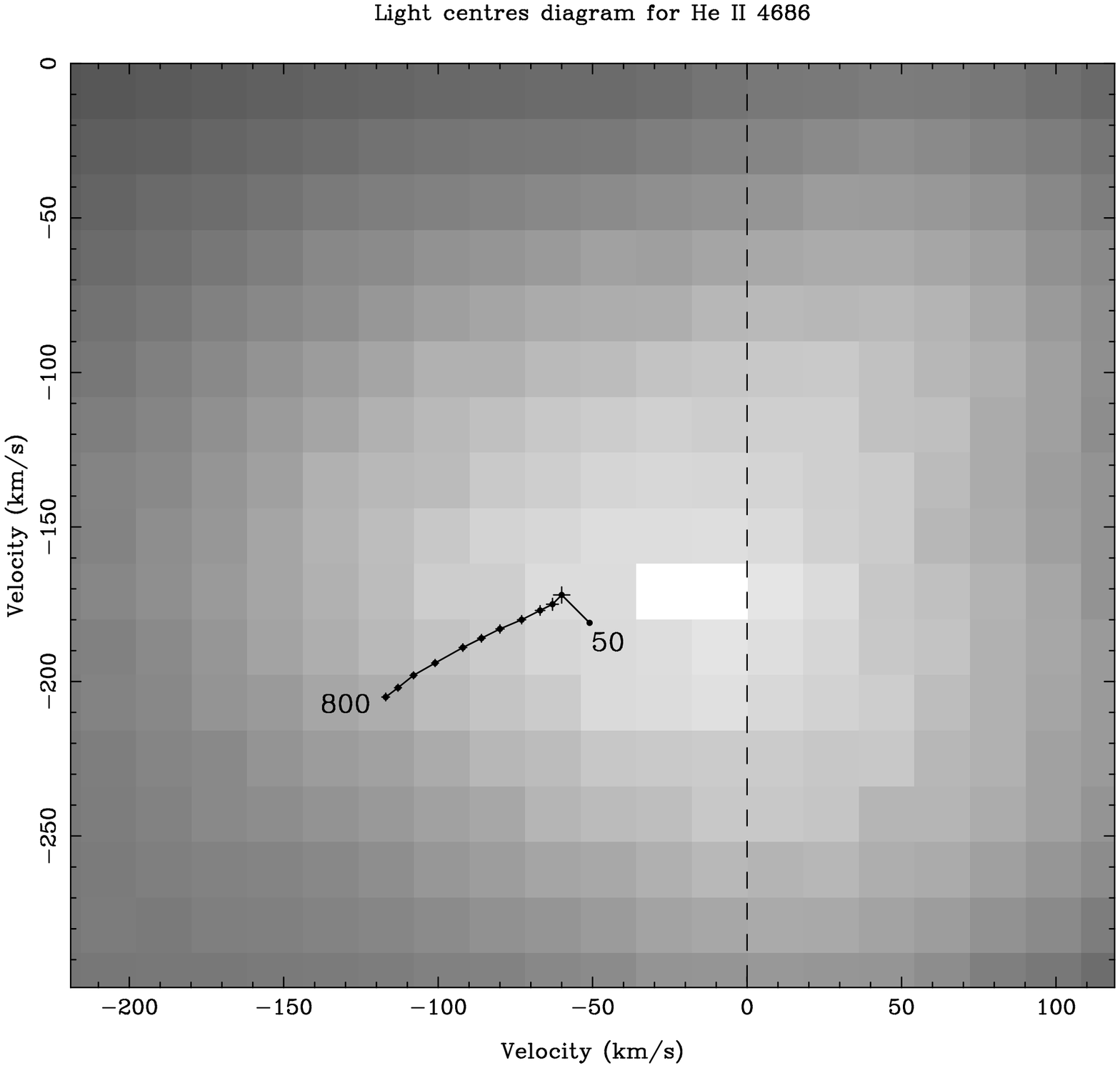,height=90mm,angle=-90}}
\caption{Light centres of the He\,{\small II} $\lambda$4686\AA\ emission line
in BT~Mon superimposed on the Doppler tomogram.
Points are plotted for radial velocity fits using single Gaussians of  
FWHM 50--800\,km\,s$^{-1}$. The dashed line represents $K_x = 0$, and is where
the white dwarf should lie.}
\label{fig:lightcentre}
\end{figure*}

\subsection{Radial velocity of the secondary star}
\label{sec:rvred}
The secondary star in BT~Mon is observable through weak absorption lines,
which are most clearly visible during eclipse. We compared regions of the 
spectra rich in absorption lines with several template red dwarfs of spectral 
types G8--K6, the spectra of which are plotted in figure~12. 
The absorption features are too weak for the normal technique of cross-correlation 
to be successful in finding the value of $K_R$, the radial velocity semi-amplitude
of the secondary star, but it is possible to exploit these features 
to obtain an estimate of $K_R$ using the technique of skew-mapping. 
This technique is described by \scite{smith93b}. 

The first step was to shift the spectra of the spectral type 
template stars to correct for
their radial velocities. We then normalised each spectrum by dividing by a spline fit
to the continuum and then subtracting 1 to set the continuum to zero.
The BT~Mon spectra were also normalised in the same way. The template spectra were
artificially broadened by 25\,km\,s$^{-1}$
to account for orbital smearing of the BT~Mon spectra through the 600-s exposures 
and then by the best-fit value of the rotational velocity of the secondary 
star found in section 4.10, $v \sin i = 138$\,km\,s$^{-1}$.  
Regions of the spectrum devoid of emission lines ($\lambda\lambda5160-5340$\AA\ and
$\lambda\lambda6400-6520$\AA) were then cross-correlated with 
each of the templates yielding a time series of cross-correlation functions (CCFs)
for each template star.

To produce the skew maps, these CCFs were back projected in the same way as 
time-resolved spectra in standard Doppler tomography \cite{marsh88b}. 
This is equivalent to
calculating line integrals along paths with parameter values 
$0<K<1000$\,km\,s$^{-1}$,
and $0<\phi_0<1$, and then plotting the line integral values in velocity space with
$K$ as the radial co-ordinate, and $\phi_0$ as the polar angle. If there is a detectable 
secondary star we would expect a peak at (0,$K_R$) in the skew maps. 
This can be repeated for each of the 
templates, and the final skew-map is the one that gives the strongest peak.

When we first back-projected the CCFs, the peak 
in each skew map was seen to be displaced to 
the left by around 40\,km\,s$^{-1}$. The reason was that we had assumed that the 
centre of mass of the system was at rest.
The systemic velocity of BT~Mon was estimated from the diagnostic 
diagram to be $\gamma = 40$\,km\,s$^{-1}$.
Applying this $\gamma$ velocity shifts the peak back towards the right of the skew map towards $K_x = 0$,
without significantly altering $K_y$, i.e. the value of $K_R$ obtained, 
\[
K_R = (K_x^2 + K_y^2)^{1/2} \approx K_y, \ K_x^2 \ll K_y^2 
\]
is almost independent of the $\gamma$ assumed. This can be understood 
by the fact that the cross-correlation peaks are strongest at phase 0, where the eclipse 
spectra lie. Changing $\gamma$ shifts the peak in the $K_x$ direction but not the $K_y$
direction. 
The positions of the peaks in the skew maps are plotted in figure~11 
against the spectral type of the template star, for $\gamma = 0$\,km\,s$^{-1}$, and
$\gamma = 40$\,km\,s$^{-1}$.

The skew maps produced using each of the template stars show well-defined peaks at
$K_y \approx 205$\,km\,s$^{-1}$, in both the blue and red. 
The final skew maps for the G8\,V template (found to be the best fitting template 
for the secondary star in section~\ref{sec:rotvred}) are shown in figure~10.
From the scatter of the points in figure~11, which correspond to the positions 
of the peaks in each skew map, we adopt a value for the radial velocity 
semi-amplitude of the secondary star from the skew maps of $K_R = 205\pm5$\,km\,s$^{-1}$. 

\begin{figure*}
\centerline{\psfig{file=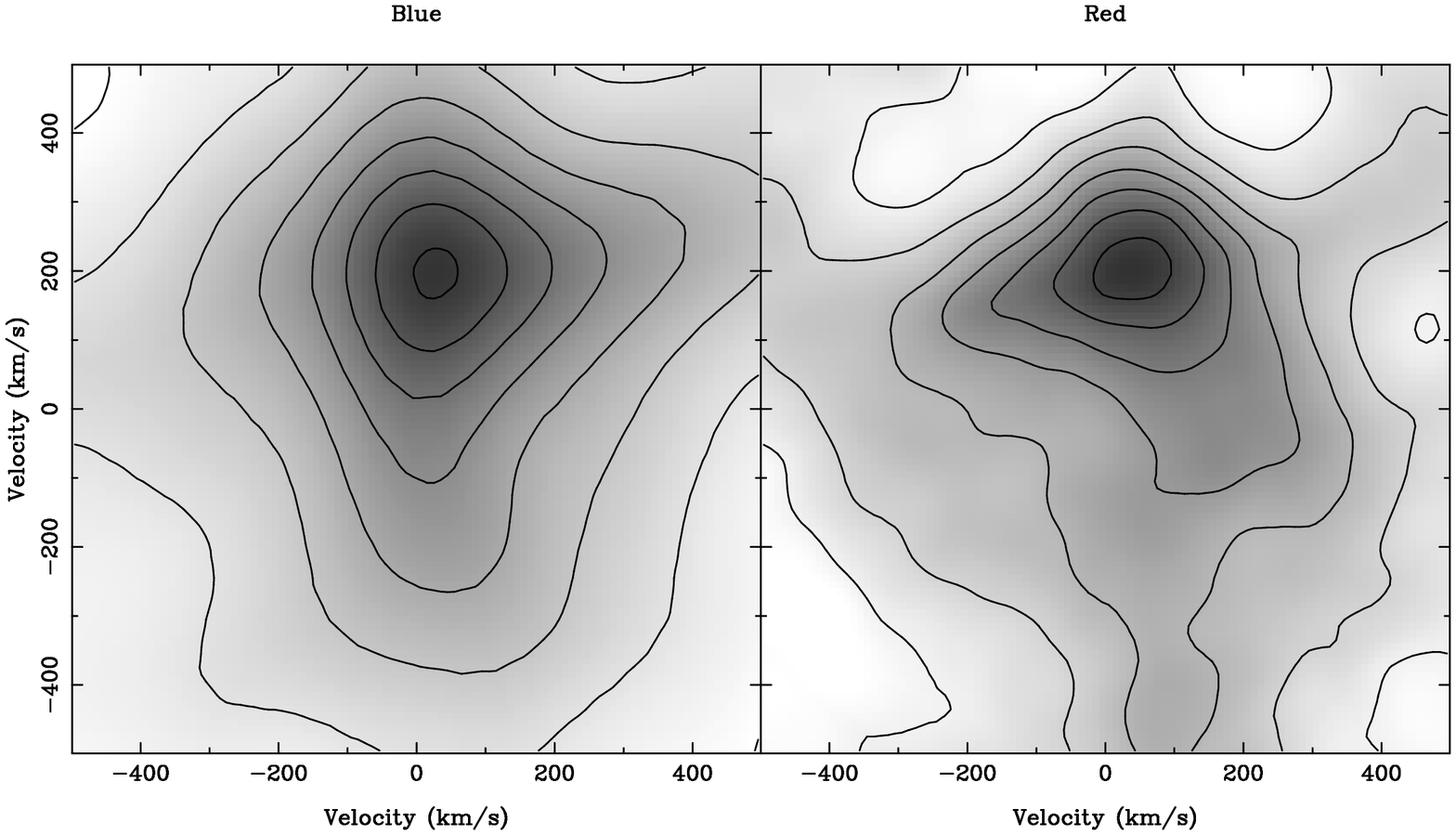,height=80mm,angle=-90}}
\caption{Skew maps of BT~Mon computed by cross-correlation with the G8\,V template.
The cross correlations were carried out over the range $\lambda\lambda$5160--5340\AA\ 
in the blue (left panel), and $\lambda\lambda$6400--6520\AA\ in the red (right panel). 
A systemic velocity of 40\,km\,s$^{-1}$ has been assumed 
in each case. The value of $K_R$ is given by the ($K_x,K_y$) value at the peak;
$K_R^2 = K_x^2 + K_y^2$}.
\label{fig:skewmap}
\end{figure*}

\begin{figure*}
\centerline{\psfig{file=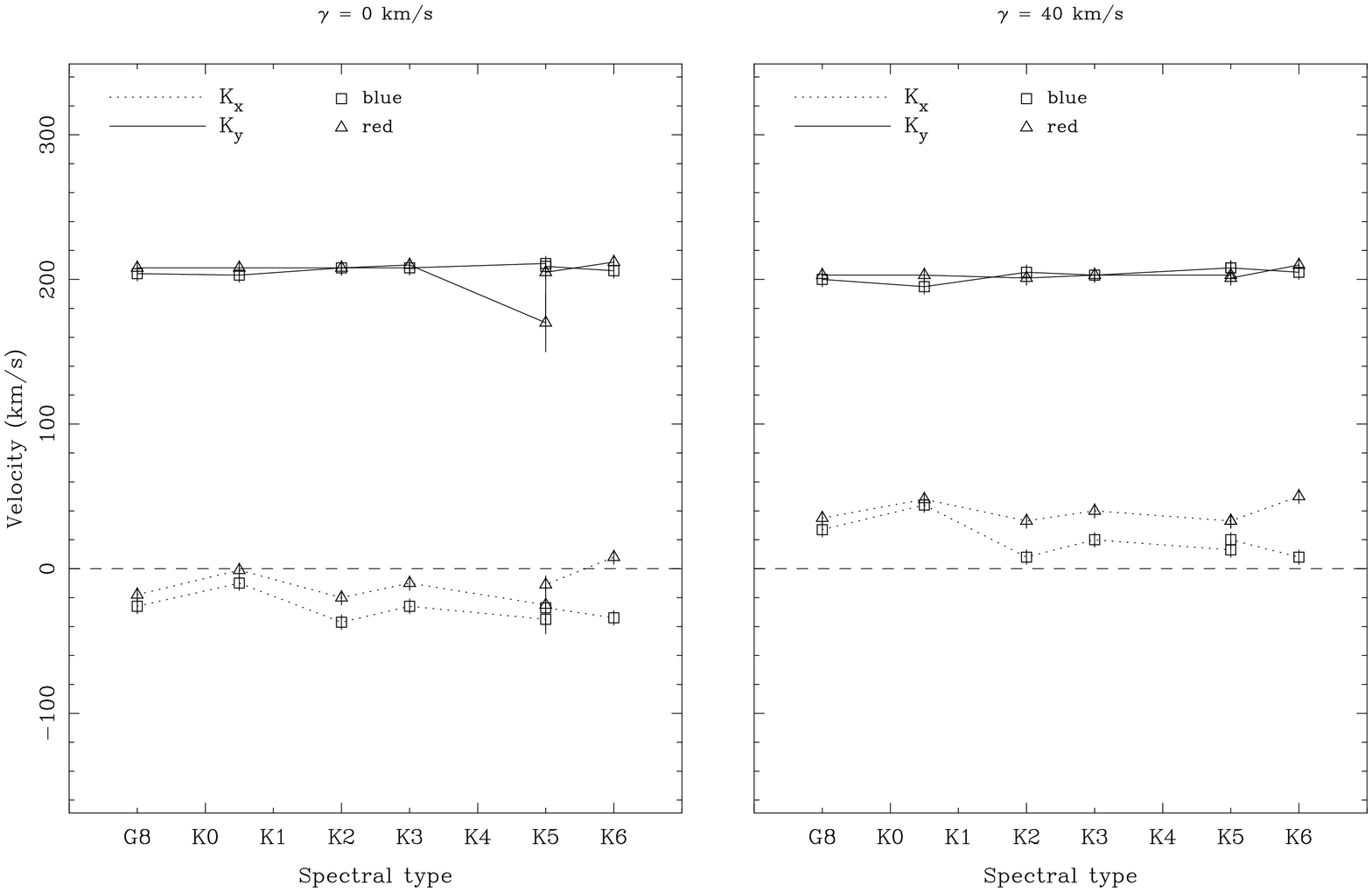,height=100mm,angle=-90}}
\caption{The position of the strongest peak in the 
skew maps, using $\gamma = 0$\,km\,s$^{-1}$ (left-hand panel), 
and $\gamma = 40$\,km\,s$^{-1}$ (right-hand panel) plotted against template 
spectral type. The skew maps produced by 
cross-correlation with the blue spectra ($\lambda\lambda$5160--5340\AA) 
are represented by squares, and those produced by cross-correlation with the 
red spectra ($\lambda\lambda$6400--6520\AA) are
represented by triangles. The dotted lines link the $K_x$ values and the solid 
lines link the $K_y$ values.}
\label{fig:krplot}
\end{figure*}

\subsection{Rotational velocity of the secondary star}
\label{sec:rotvred}
Having obtained estimates of $K_W$ and $K_R$, the
third measurement we made was the rotational velocity of the secondary
star, $v \sin i$, which was found in the following way.
Using the value of $K_R$ obtained from the skew maps in section~4.9 
we corrected for the orbital motion 
when averaging the four mid-eclipse spectra of BT~Mon (those which show the 
strongest absorption features). We then broadened the template spectra to account for
smearing due to the orbital motion of BT~Mon over each 600-s exposure -- about 
25\,km\,s$^{-1}$ -- and then rotationally broadened the templates by a range of 
velocities (50--200\,km\,s$^{-1}$). Finally we ran an optimal subtraction routine, which
subtracts a constant times the normalised template spectra from the 
normalised, orbitally-corrected BT~Mon eclipse spectrum,
adjusting the constant to minimise the residual scatter between the spectra.
The scatter is measured by carrying out the subtraction and then computing the   
$\chi^2$ between this residual spectrum and a smoothed version of itself.
By finding the value of the rotational broadening which minimises the reduced-$\chi^2$ 
we obtain an estimate of both $v\sin i$ and the spectral type 
of the secondary star. There should, strictly speaking, be a 
small correction due to the intrinsic rotational velocity of the template star, but
as the templates are late type stars they are assumed to be 
sufficiently old to have lost most of their angular momentum by magnetic braking
and to have a very small
$v\sin i$ (of the order of 1\,km\,s$^{-1}$; \pcite{gray92}). 

The value of $v\sin i$ obtained with this method depends upon several factors.
Using later spectral types as templates tended to produce lower values of $v\sin i$
by up to 25\,km\,s$^{-1}$ in the blue band, but had a smaller effect in the red
with a spread of only 6\,km\,s$^{-1}$. 
The wavelength range we used for the optimal subtraction 
also affected the value, 
as did the value of the limb-darkening coefficient used in the broadening 
procedure, 
and the amount of smoothing of the residual spectrum in the 
calculation of $\chi^2$ in the optimal subtraction routine.
The values of $v\sin i$ for the different templates in the red and the blue wavelength ranges,
calculated using values for the limb-darkening coefficient of 0.5 and the smoothing coefficient 
of 15\,km\,s$^{-1}$ are listed in table~3. 

We plotted the values of $\chi^2$ versus $v\sin i$ for each of the spectral type
template stars in figure~13. The minimum of the lowest curve 
gives both the value of $v\sin i$ and the spectral type of the secondary star.
The two wavelength ranges used give very consistent values. The blue band 
gives $v\sin i \approx 140$\,km\,s$^{-1}$, and the spectral type G8\,V, 
or possibly earlier, since we have no earlier template spectra.  
The red band also gives a spectral type of the secondary star of G8\,V or earlier, 
although note the expanded vertical scale in the right hand panel, 
and a $v\sin i \approx 140$\,km\,s$^{-1}$.
We adopt a value of $v\sin i = 138\pm5~$\,km\,s$^{-1}$ for the rotational velocity 
of the secondary star, with the notional errors 
on this estimate reflecting all of the variations noted in the previous paragraph.
The spectral type of the secondary star is found to be G8\,V from these $v\sin i$ curves and 
also from a visual comparison of the absorption line depths, the $V-R$ colour found in 
section~\ref{sec:distance}, and the mass and radius found in section~\ref{sec:parameters}. 

The optimal subtraction technique also tells us the value of the constant by
which the template spectra were multiplied, which, for normalised spectra, 
is the fractional contribution of the secondary star to the total light. 
The results are plotted in figure~14, both in comparison with the eclipse and 
non-eclipse spectra. We find that the secondary star contributes 
$85 \pm 8$ per cent of the total blue light in eclipse, and
$81 \pm 35$ per cent of the total red light in eclipse.

\begin{figure*}
\centerline{\psfig{file=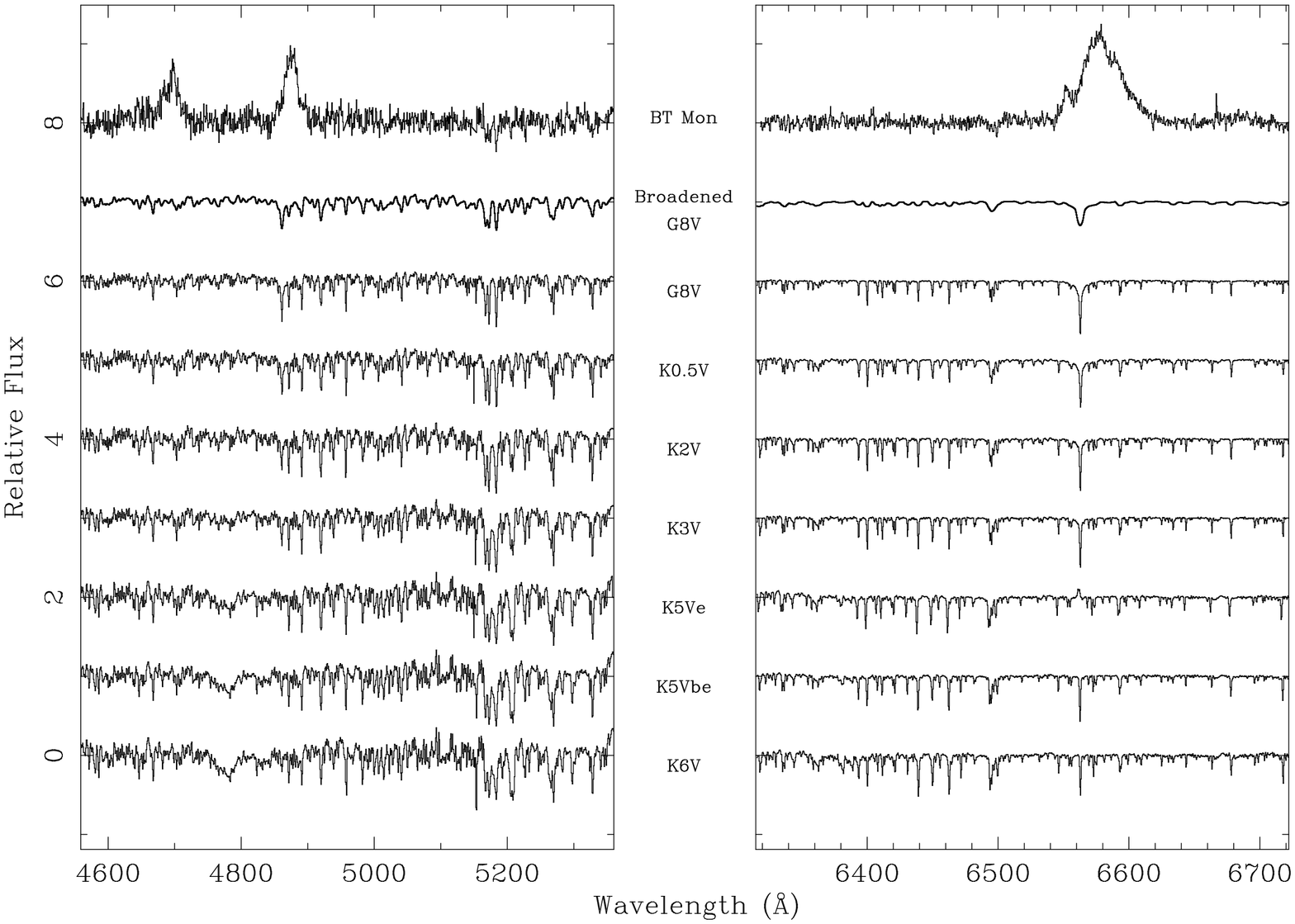,height=4.0in,angle=-90}}
\caption{Normalised spectra of the seven template stars. The offset between each is 1.
The uppermost spectra are a normalised average of the four mid-eclipse spectra of BT~Mon,
when its absorption features are most visible. The spectrum below that of BT~Mon is that
of the best fitting template star (G8\,V) which has been broadened by 25\,km\,s$^{-1}$ 
to account for orbital smearing of the BT~Mon spectra during exposure and by 138\,km\,s$^{-1}$
to account for the rotational broadening of the lines in BT~Mon.}
\label{fig:spectra}
\end{figure*}

\begin{figure*}
\centerline{\psfig{file=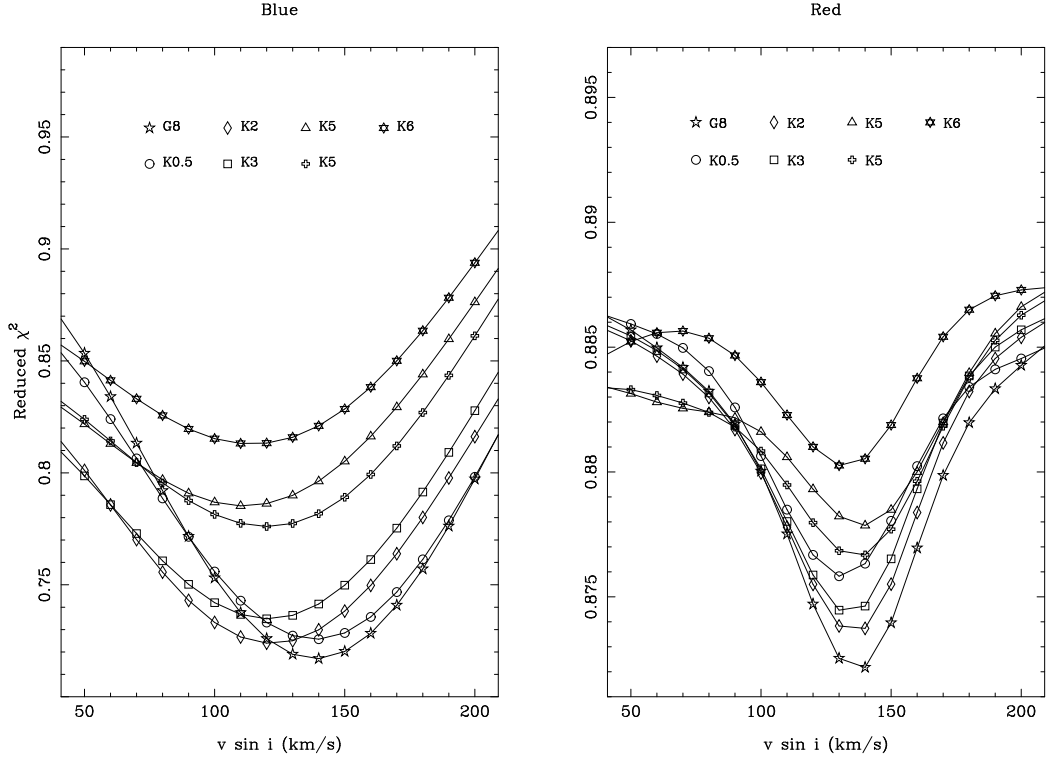,height=100mm,angle=-90}}
\caption{Reduced-$\chi^2$ from the optimal subtraction technique plotted against 
$v\sin i$ for both the blue wavelength range ($\lambda\lambda$5160--5340\AA, left-hand panel)
and the red wavelength range ($\lambda\lambda$6400--6520\AA, right-hand panel). 
The red band has 304 degrees of freedom, and the blue 216 degrees of freedom.}
\label{fig:vsini}
\end{figure*}

\begin{figure*}
\centerline{\psfig{file=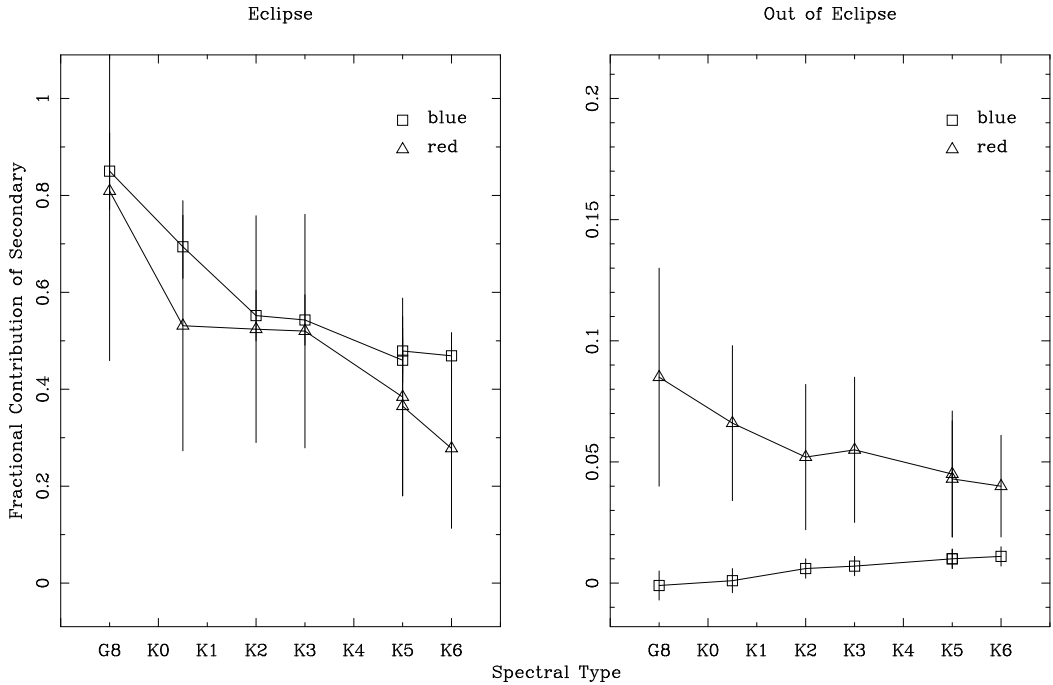,height=90mm,angle=-90}}
\caption{The fractional contribution of the secondary star to the total light
in eclipse (left-hand panel) and out of eclipse (right-hand panel) in the
red (plotted as triangles) and the blue (plotted as squares), 
as determined by the optimal subtraction of the template spectra.}
\label{fig:contrib}
\end{figure*}

\begin{table*}
\caption{Values of $v\sin i$ for BT~Mon as measured 
by comparison with the rotationally broadened profiles of G8--K6 
dwarf templates. Also listed are the fractional contributions of the secondary 
star to the total light during eclipse, and the position 
	of the strongest peak in the skew maps for $\gamma=40$\,km\,s$^{-1}$.} 
\boldmath{
{\normalsize\bf
\begin{tabular}{lcccccc} 
 & & & & & & \\
\multicolumn{1}{l}{Template} &
\multicolumn{1}{c}{$v \sin i$} & 
\multicolumn{1}{c}{$v \sin i$} & 
\multicolumn{1}{c}{Fractional} & 
\multicolumn{1}{c}{Fractional} & 
\multicolumn{1}{c}{($K_x$,$K_y$) from} & 
\multicolumn{1}{c}{($K_x$,$K_y$) from} \\
 & at min $\chi^2$ & at min $\chi^2$ & contribution & contribution & 
blue skew map & red skew map \\
 & (blue) & (red) & of secondary & of secondary & $\gamma=40$\,km\,s$^{-1}$ & $\gamma=40$\,km\,s$^{-1}$ \\
 & km\,s$^{-1}$ & km\,s$^{-1}$ & (blue) & (red) & (km\,s$^{-1})$ & (km\,s$^{-1}$) \\
 & & & & & & \\
G8V & 139 & 137 & $0.85\pm0.08$ & $0.81\pm0.35$ & (27,200) & (35,203) \\
K0.5V & 139 & 132 & $0.69\pm0.07$ & $0.53\pm0.26$ & (44,195) & (48,203) \\
K2V & 122 & 136 & $0.55\pm0.05$ & $0.52\pm0.23$ & \ (8,205) & (33,201) \\
K3V & 121 & 134 & $0.54\pm0.05$ & $0.52\pm0.24$ & (20,203) & (40,203) \\
K5Ve & 112 & 139 & $0.46\pm0.05$ & $0.38\pm0.20$ & (13,208) & (33,203) \\
K5Vbe & 121 & 137 & $0.48\pm0.05$ & $0.37\pm0.19$ & (20,208) & (33,201) \\
K6V & 115 & 133 & $0.47\pm0.05$ & $0.28\pm0.17$ & \ (8,205) & (50,210) \\
 & & & & & & \\
\end{tabular}
}
}
\label{tab:templates}
\end{table*}
\rm

\subsection{The distance to BT~Mon}
\label{sec:distance}
By finding the apparent magnitude of the secondary star from its contribution to the total 
light during eclipse, and 
estimating its absolute magnitude, we can calculate 
the distance to BT~Mon. At mid-eclipse, the apparent magnitude of the system 
is $17.4\pm0.2$ in the red wavelength range, which is approximately the R band, 
of which the secondary star contributes $81\pm35$\,per cent. In the blue (approximately 
the V band), the apparent magnitude is $18.0\pm0.2$ in eclipse, 
of which the secondary star contributes $85\pm8$\,per cent. 
This gives us rough apparent magnitudes of $R=17.6\pm0.5$ and $V=18.2\pm0.3$
for the secondary star.
There are a number of ways of estimating the absolute magnitude of the 
secondary star (\pcite{serrano78}, \pcite{patterson84}, \pcite{warner95b}, \pcite{gray92}),
assuming it is on the main sequence. 
We took each of these into account and 
adopted an averaged value of the absolute magnitude of $M_V=+6.0\pm0.5$ 
for the secondary star and hence $m_V - M_V = 12.2 \pm 0.5$.
\scite{duerbeck81} calculated the extinction to BT~Mon to be $A_V = 0.63$ magnitudes 
per kpc (we assume the error is small).
The distance to BT~Mon can then be calculated using the equation
\begin{equation}
5\log(d/10) = m_V - M_V - d\,A_V/1000
\end{equation}
and is found to be $1700\pm300$\,pc.

Another method of finding the distance is to determine the angular diameter of the secondary 
star from the observed flux and a surface brightness calibration that we derive 
from the Barnes-Evans relation \cite{barnes76}.
\begin{equation}
F_v = 4.2207 - 0.1 V_0 - 0.5 \log \phi = 3.977 - 0.429(V-R)_0
\end{equation}
where $V_0$ and $(V-R)_0$ are the unreddened V magnitude and (V-R) colour
of the secondary star.
We take the unreddened colour to be $(V-R)_0 = 0.6 \pm0.1$ (typical of a late G or early K 
dwarf) and $V_0 = 17.2 \pm 0.3$ (estimating the extinction to be 1.0 magnitudes, from above).
Using these values and the radius of the secondary star derived in section~\ref{sec:parameters}
we obtain a distance of $2300\pm700$\,pc.

A number of authors have estimated the distance to BT~Mon. \scite{sanford40} 
derived a value of 700--900\,pc from the strengths of the interstellar 
H and K lines; \scite{mclaughlin41} assumed the radial velocities of the 
interstellar lines were due to the differential rotation of the Galaxy, 
and by also assuming that
calcium was uniformly distributed along the line of sight, was able to
estimate the distance to BT~Mon. Using his method and the modern values for
Oort's constants, RNK found a distance of 1400\,pc to BT~Mon. They further
constrained the distance to be 1000--2000\,pc by combining the value of the
reddening given by \scite{wade81} with the reddening 
versus distance curves of \scite{deutschmann76}.
\scite{marsh83} estimated the distance to be $\sim 1800$\,pc
from the expansion velocity of the nebula. 

\subsection{System parameters}
\label{sec:parameters}
Our measurements of 
the radial velocity of the secondary star, $K_R = 205\pm5$\,km\,s$^{-1}$ and the
rotational broadening, $v\sin i = 138\pm5$\,km\,s$^{-1}$, as well as our
measurement of the radial velocity of the primary star $K_W = 170\pm10$\,km\,s$^{-1}$ 
can now be used in conjunction with our newly derived period and RNK's measurement
of the eclipse full width at half depth,
$\Delta\phi_{1/2} = 0.117\pm0.011$, to determine accurate system parameters for BT~Mon.
Because only four of these five measurements are needed to calculate the system parameters,
and because as they stand, the measured values of these parameters are not
consistent with one another, we have opted for a Monte Carlo approach similar to that of 
Horne et al. (1993) to calculate the masses and other parameters. For a given set of values of
$K_R$, $v\sin i$ , $\Delta\phi_{1/2}$ and $P$ the other system parameters are calculated 
as follows.

Because the secondary star fills its Roche lobe, ${R_2}/a$ can be estimated; 
$R_2$ is the equatorial radius of the secondary star and $a$ is the binary 
separation. 
We used Eggleton's formula \cite{eggleton83} which gives the volume-equivalent radius
of the Roche lobe to better than 1\%, which is close to the equatorial radius of 
the secondary star as seen during eclipse; 
\begin{equation}
{{R_2} \over a} = {{0.49q^{2/3}} \over {{0.6q^{2/3} + \ln{(1+q^{1/3})}}}}. 
\end{equation}
The secondary star rotates synchronously with the orbital motion, so we can
combine $K_R$ and $v\sin i$, to get
\begin{equation}
{{R_2} \over a}(1+q) = {{v\sin i} \over {K_R}}.
\end{equation}
This gives us two simultaneous equations so we can calculate the mass ratio $q$
and $R_2/a$.

Simple geometric considerations give us
\begin{equation}
\Bigl({{R_2} \over a}\Bigr)^2 = \sin^2\pi\Delta\phi_{1/2}+\cos^2\pi\Delta\phi_{1/2}\cos^2i.
\end{equation}
which using the value of $R_2/a$ obtained using equations (5) and (6) allows us to calculate
the inclination of the system.
Kepler's Third Law gives us
\begin{equation}
{{K_R^3P}\over{2\pi G}}={{M_1\sin^3i}\over{(1+q)}^2}
\end{equation}
which with the values of $q$ and $i$ calculated using equations (5), (6) and (7), 
gives the mass of the primary star. The mass of the secondary star can then be obtained using 
\begin{equation}
M_2=qM_1.
\end{equation}
The radius of the secondary star is obtained from the equation
\begin{equation}
{{v\sin i} \over {R_2}} = {2\pi\sin i \over P}, 
\end{equation}
and the separation of the components, $a$ is calculated from equations (6) and (10)
with $q$ and $i$ now known.

The Monte Carlo simulation takes 10,000 sample values of $K_R$, $v\sin i$ and 
$\Delta\phi_{1/2}$, treating each as being normally distributed about their measured 
values with standard deviations equal to the errors on the measurements. 
We then calculate the masses of the components, the inclination of the system, 
the radius of the secondary star, and the 
separation of the components, as outlined above, omitting $(K_R, v\sin i, \Delta\phi_{1/2})$
triplets which are inconsistent with $\sin i \leq 1$.
Each triplet is also subject to rejection if the calculated value of $K_W$ (implied
by the triplet) is inconsistent with the measured value of $K_W$ (implied by figure~9).
This was done in accordance with a Gaussian probability law, i.e. 
if the calculated value of $K_W$ lies $1\sigma$ (with $\sigma$ equal to the measured error 
on $K_W$) from the mean (the measured value of $K_W$), 
the probability of rejection is 68 per cent, at $2\sigma$ it is 95 per cent, etc.
Each accepted $(M_1,M_2)$ pair is then plotted as a point in figure~15,  
and the masses and their errors are computed from the mean and standard deviation of 
the distribution of spots. The solid curves in figure~15 satisfy the white dwarf radial 
velocity constraint, 
$K_W = 170\pm10$\,km\,s$^{-1}$, the secondary star radial velocity constraint 
$K_R = 205\pm5$\,km\,s$^{-1}$ and the rotational velocity of the secondary star,
$v \sin i = 138\pm5$\,km\,s$^{-1}$. 
We find that $M_1 = 1.04 \pm 0.06$M$_\odot$ and 
$M_2 = 0.87 \pm 0.06$M$_\odot$
The values of all the system parameters deduced from 
the Monte-Carlo computation are listed in table~4. 

The mass and radius of the secondary star are consistent within the error bars
with those of a main sequence star according to the values published by \scite{gray92} 
for a G8 dwarf ($M = 0.89$M$_\odot$, $R = 0.88$R$_\odot$) and
the empirical relation obtained by \cite{warner95b}
between mass and radius for detached stars in binaries and the secondary stars in CVs,
\begin{equation}
R/R_\odot=(M/M_\odot)^{13/15}.
\end{equation}
The secondary star in BT~Mon appears to be a main sequence star in all respects. 

\begin{figure*}
\centerline{\psfig{file=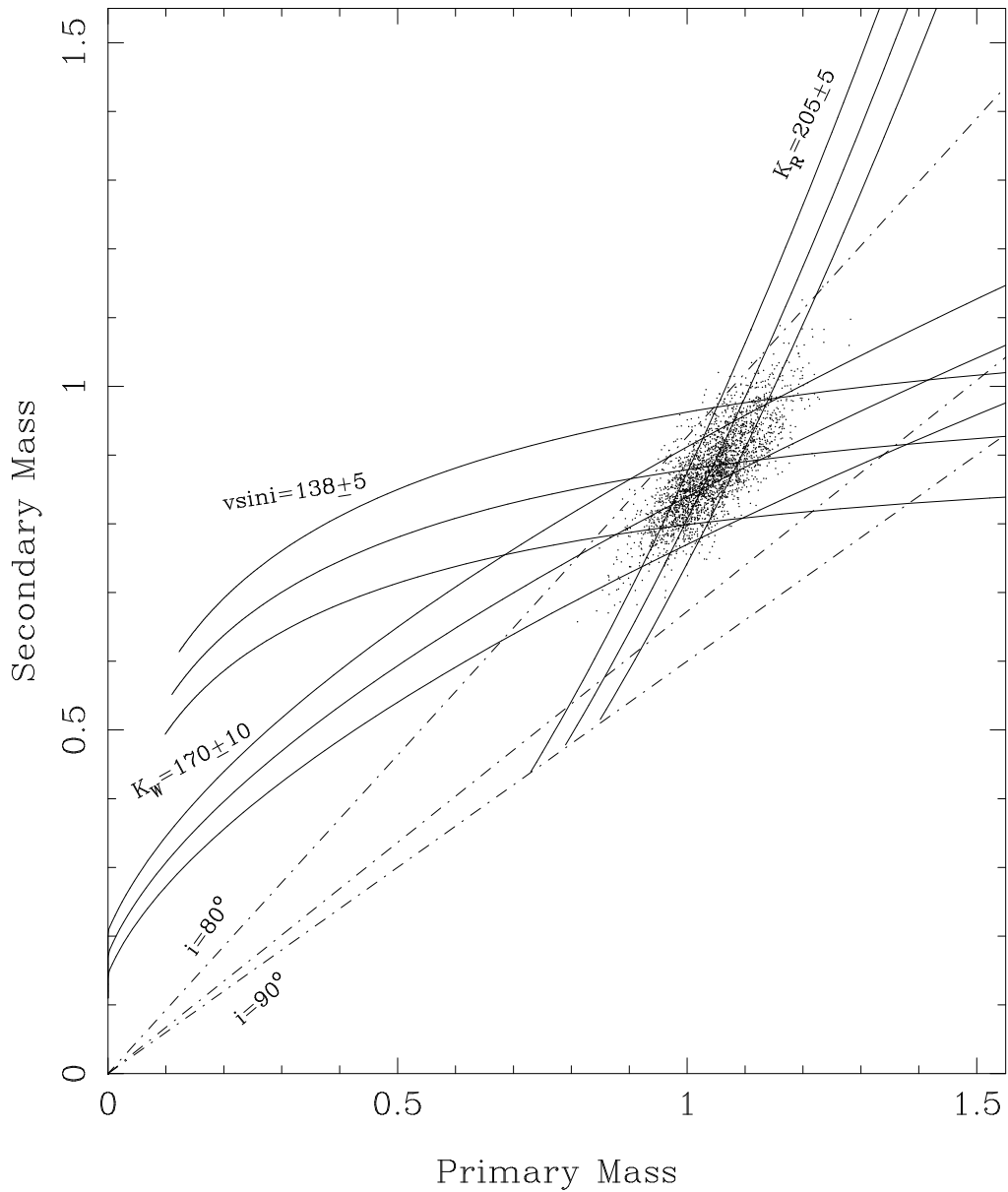,height=160mm,angle=0}}
\caption{Constraints on the masses of the stars in BT~Mon. Each dot represents an 
($M_1,M_2$) pair. Dashed-dotted lines 
are lines of constant inclination (i=80\degr, i=85\degr, i=90\degr). The
solid curves satisfy the constraints from the secondary star radial velocity 
$K_R = 205\pm5$\,km\,s$^{-1}$, the primary star radial velocity 
$K_W = 170\pm10$\,km\,s$^{-1}$ and the rotational velocity of the secondary star
$v \sin i = 138\pm5$\,km\,s$^{-1}$.}
\label{fig:monte}
\end{figure*}

\begin{table}
\caption{System Parameters for BT Mon}
\boldmath{
{\normalsize\bf
\begin{tabular}{lcc} 
 & & \\
\multicolumn{1}{l}{Parameter} &
\multicolumn{1}{c}{Measured} & 
\multicolumn{1}{c}{Monte Carlo} \\
\multicolumn{1}{c}{ } & 
\multicolumn{1}{c}{Values} &
\multicolumn{1}{c}{Values} \\
 & & \\
$P_{orb}$ (d) & 0.33381379 & \\
$K_R$ (km\,s$^{-1}$) & $205\pm5$ & $205\pm5$ \\
$v$\,sin $i$ (km\,s$^{-1}$) & $138\pm5$ & $136\pm4$ \\
$\Delta\phi_{1/2}$ & $0.117\pm0.011$ & $0.109\pm0.007$ \\
$K_W$ (km\,s$^{-1}$) &$ 170\pm10$ & $171\pm8$ \\
$q$ & & $0.84\pm0.04$ \\
$i\degr$ & & $82.2\pm3.2$ \\
$M_1/M_\odot$	& & $1.04\pm0.06$ \\
$M_2/M_\odot$ & & $0.87\pm0.06$ \\
$R_2/R_\odot$ & & $0.89\pm0.02$ \\
$a/R_\odot$ & & $2.46\pm0.05$ \\
Distance (pc) & $1700\pm300$ & \\
Spectral type & G8\,V & \\
of secondary  & & \\
 & & \\
\end{tabular}
}
}
\label{tab:parameters}
\end{table}

\section{Discussion}
\label{sec:discuss}

\subsection{The nature of the system}
\label{sec:nature}
The emission lines in the spectrum of BT~Mon show single-peaked profiles,
instead of the double-peaked profiles expected of high-inclination 
accretion discs \cite{horne86}. These profiles, and the phase 0.5 absorption
in the Balmer lines are characteristics of the SW~Sex stars, a class
of the nova-like variables \cite{thorstensen91}. 

There are several competing models to explain the line emission from the
SW~Sex objects. \scite{honeycutt86} invoked the existence of an accretion
disc wind which dominates the Balmer emission and remains unobscured
during primary eclipse. 
\scite{williams89} has suggested that the
accretion disc is disrupted by a strong magnetic field due to the white 
dwarf, and accretion occurs via the field lines directly onto the white 
dwarf. The single-peaked Balmer lines would then be produced by gas flowing
in an accretion curtain lying above the orbital plane.
Dhillon et al. (1997) attribute the emission to an extended bright spot, the
secondary star and an accretion disc.
\scite{hellier94} suggests that the emission-line peculiarities of PX~And 
and other SW~Sex stars can be explained by an accretion stream which 
overflows the initial impact with the accretion disc
and continues to a later re-impact close to the white dwarf. 

The flares we have discovered in the line emission lead us to favour the 
intermediate polar interpretation of the system, in which
material is accreted onto the magnetic poles of the primary star,
giving rise to hot spots on its surface. As the white dwarf rotates
asynchronously, these hot spots illuminate the inner parts of the disc,
the accretion stream or the accretion curtain 
like a lighthouse, and we observe a flare. This is consistent with
the apparent slant to the flares; if the flares were due to blobby accretion 
or some other form of quasi-periodic flickering then the flares would presumably appear
horizontal in the trailed spectra. 
White et al. (1996) dismissed the IP model for BT~Mon, which identifies the high velocity 
emission as coming from a magnetic accretion funnel, because the high velocity
spike they observed in the radial velocity curve suggested to them that the
feature only appeared during eclipse, whereas in our data it is visible, as it should be
according to this model, throughout the whole orbital cycle. It would be of 
interest to search for the flares seen in BT~Mon in the SW~Sex stars.

The high velocity component we see in the Balmer emission of BT~Mon is 
visible throughout the orbital cycle and does not appear to be eclipsed, implying that 
the source lies above the orbital plane. It is similar to
that seen in some polar systems e.g. HU~Aqr \cite{schwope97}, in which the source
of the high velocity emission is thought to be the magnetic accretion 
funnel. An accretion funnel may be the source of the high velocity component in BT~Mon, 
or it may come from the accretion stream overflowing the (undetected) disc, as suggested by 
\scite{hellier94} as a possible mechanism for the SW~Sex stars, but our Doppler maps
are not fully consistent with BT~Mon satisfying either of these models. 

For us to be sure of the classification of BT~Mon as an intermediate polar, 
the flares need to be observed to be
coherent, and our data set has neither the time resolution nor the 
baseline to give positive evidence of coherence. The system would also
be expected to show X-ray pulsations and if it were strongly magnetic, 
circular and linear polarisation (not observed by \pcite{stockman92}).
The luminosity of BT~Mon in soft X-rays is low since it has not been 
detected by \it ROSAT \rm in a survey of classical novae
(Marina Orio, private communication).
BT~Mon lies at a low galactic latitude at a distance of $\sim2$\,kpc,
so even if it were intrinsically luminous in X-rays, it is possible that 
the high level of extinction would severely diminish its apparent brightness
in the soft X-ray band although hard X-rays should still be visible.

\subsection{The TNR model}
\label{sec:tnr}
The motivation for studying BT~Mon was to obtain a tight constraint on the
mass of the white dwarf, which we have done successfully, in order to test
the thermonuclear runaway model for novae outbursts, which broadly predicts
that the higher the mass of the white dwarf, the faster the nova outburst. 
Unfortunately the literature is divided over the speed class of BT~Mon.

Nova~Mon~1939 was not discovered until 1939~Oct~8 \cite{sanford40}, by 
which time it had declined from its maximum brightness to around 8th or 9th 
magnitude, its rise to maximum failing to be observed because this occurred 
when it was in the daytime sky. \scite{schaefer83} looked back over archival photographic 
plates, taken in the month prior to the first observations by \scite{whipple40}, 
which they believed to show the nova at maximum light, with a  
magnitude of 8.5 (although \pcite{duerbeck87b} notes an observation 
which gives the magnitude of BT~Mon as 7.6), and combined 
these with later observations to produce an eruption light curve which
suggests BT~Mon was a slow nova, giving $t_2 = 140$ days, and 
$t_3 = 190$ days, where $t_2$ and $t_3$ are the times for the nova to
decline by 2 and 3 magnitudes from its maximum brightness. 

This contradicts the earlier work of \scite{mclaughlin41} who compared the 
spectra of 1939~Dec~26 and 1940~Jan~17
to those of other novae. Assuming normal behaviour, the spectra of BT~Mon 
corresponded to declines of 5.4 magnitudes and 5.9 magnitudes, respectively, from maximum.
The observed apparent magnitudes were 9.6 and 10.0 \cite{whipple40}
which tells us that the faintest it could have been at maximum brightness was
an apparent magnitude of 4.2. McLaughlin extrapolated the 
light curve back, and found that it corresponded to that of a fast nova; 
he later published a value for $t_3$ of 36 days \cite{mclaughlin45}.

One possible resolution of this problem is to image the nebula. Fast novae
generally give rise to almost spherical shells, slow novae to more aspherical shells
(\pcite{slavin95}; \pcite{lloyd97}). An image of the shell of BT~Mon
has been obtained by \scite{duerbeck87a}, who found it to be very weak
with a nearly circular outline, which makes us favour the interpretation of
BT~Mon as a fast nova. 

\section{Conclusions}
We have shown that BT~Mon has a white dwarf with a high mass ($1.04\pm0.06M_\odot$)
and that the observational evidence supports the classification of BT~Mon as a
fast nova, in keeping with the thermonuclear runaway model of nova outbursts.
The calculated mass and radius, and the spectral features of the secondary star, 
are all consistent with it being a G8 main sequence star.
We have also found weak, regular flaring in the line emission, which could
indicate that the system is an intermediate polar. 

\section*{\sc Acknowledgements}

We are indebted to Marina Orio for the communication of the results of her 
BT~Mon ROSAT observations.
DAS is supported by a PPARC studentship and TRM by a PPARC advanced fellowship.
The WHT is operated on the island of La Palma by the Royal Greenwich Observatory 
in the Spanish Observatorio del Roque de los Muchachos of the IAC.

\bibliographystyle{mnras}
\bibliography{btmon}

\end{document}